\documentclass[a4,10pt]{article}

\usepackage{epsfig}
\usepackage{dcolumn}
\usepackage{amsmath}
\usepackage{color}
\usepackage{graphicx}
\usepackage{amsfonts}
\def \be  {\begin{equation}}
\def \ee  {\end{equation}}
\def \ee  {\end{equation}}
\def \bea {\begin{eqnarray}}
\def \eea {\end{eqnarray}}

\newcommand{\ba} {\begin{eqnarray}}
\newcommand{\ea} {\end{eqnarray}}
\usepackage{graphicx}% Include figure files
\usepackage{dcolumn} % Align table columns on decimal point
\usepackage{bm}      % bold math

\begin{document}

%\preprint{ECTP-2016-05}
%\preprint{WLCAPP-2016-05}
%\vspace*{3mm}

\title{Chemical freezeout parameters within generic nonextensive statistics}

\author{Abdel Nasser TAWFIK\footnote{Corresponding author: a.tawfik@eng.mti.edu.eg}\\
{\footnotesize Egyptian Center for Theoretical Physics (ECTP), Modern University for Technology and  Information (MTI), 11571 Cairo, Egypt}\\
{\footnotesize World Laboratory for Cosmology And Particle Physics (WLCAPP), 11571 Cairo, Egypt}\\
\\
{Hayam Yassin and Eman R. Abo Elyazeed}\\
{\footnotesize Physics Department, Faculty of Women for Arts, Science and Education, Ain Shams University, 11577 Cairo, Egypt}
}

\date{}
\maketitle

\begin{abstract}

The particle production in relativistic heavy-ion collisions seems to be created in a dynamically disordered system which can be best described by an extended exponential entropy. In distinguishing between the applicability of this and Boltzmann-Gibbs (BG) in generating various particle-ratios, generic (non)extensive statistics (GNS) is introduced to the hadron resonance gas model. Accordingly, the degree of (non)extensivity is determined by the possible modifications in the phase space. Both BG extensivity and Tsallis nonextensivity are included as very special cases defined by specific values of the equivalence classes $(c, d)$. We found that the particle ratios at energies ranging between $3.8$ and $2760~$GeV are best reproduced by nonextensive statistics, where $c$ and $d$ range between $\sim0.9$ and $\sim1$. 
The present work aims at illustrating that the proposed approach is well capable to manifest the statistical nature of the system on interest. We don't aim at highlighting deeper physical insights. In other words, while the resulting nonextensivity is neither BG nor Tsallis, the freezeout parameters are found very compatible with BG and accordingly with the well-known freezeout phase-diagram, which is in an excellent agreement with recent lattice calculations. We conclude that the particle production is nonextensive but should not necessarily be accompanied by a radical change in the intensive or extensive thermodynamic quantities, such as internal energy and temperature. Only, the two critical exponents defining the equivalence classes $(c, d)$ are the physical parameters characterizing the (non)extensivity.

\end{abstract}


\section{Introduction}
\label{sec:intro}

A comprehensive description for various phenomenon in different high-energy experiments \cite{Tawfik:2014eba,Tawfik:2010aq} can be given by extensive [such as Boltzmann-Gibbs (BG)] and nonextensive (such as Tsallis) statistics. In both statistical approaches, the thermodynamic consistency should be guaranteed. In proving that the fireballs or heavy resonances lead to a bootstrap approach, i.e. further fireballs consist of smaller fireballs and so on, extensive statistics was utilized by Hagedorn \cite{Hagedorn1965}. Assuming that the characteristic distribution-function gets variations from a possible symmetrical change, etc., nonextensive concept was first introduced for the particle productions \cite{Tawfik:2010uh}. The implementation of nonextensive Tsallis statistics was first introduced in Refs. \cite{ReFF1,ReFF2,Tsallis:1987eu,Prato:1999jj}, with a clear emphasize that the phase space plays an essential role. It has been argued that substituting the Boltzmann factor by $q$-exponential function with $q>1$ leads to a good agreement with the experimental results at high energies, especially the transverse momentum spectra. Is is widely accepted that the statistical fits of the the transverse momentum spectra characterize the kinetic freezeout, which is conjectured  to take place later (in the sense that the system cools down) after the chemical freezeout \cite{Tawfik:2014eba}. Recently, Tawfik explained that this procedure is not necessarily fully incorporating the nonextensivity in the particle productions, even at the kinetic freezeout \cite{Tawfik:2016pwz,Tawfik:2016jol,Tawfik:2017bul,Cleymans:2011in,Azmi:2014dwa}. The long-range fluctuations, the correlations, and the interactions besides the possible modifications in the phase space of the particle production are not properly incorporated through Tsallis algebra. In long-range interactions, both thermodynamic and long-time limits do not commute. Therefore, generic nonextensive statistics (GNS) was introduced in Ref. \cite{Tawfik:2016pwz,Tawfik:2016jol,Tawfik:2017bul}, in which the phase space becomes responsible in determining the degree of (non)extensivity. It was shown that the lattice thermodynamics is well reproduced when the proposed GNS become characterized by extensive critical exponents ($1, 1$), while the heavy-ion particle ratios are only reproduced when the proposed GNS become nonextensive critical exponents, e.g. neither $0$ nor $1$. The latter differs from Tsallis \cite{Tawfik:2016pwz,Tawfik:2016jol,Tawfik:2017bul}.

Nonextensive statistics becomes the relevant approach for nonequilibrium stationary states. While zeroth law of thermodynamics in equilibrium introduces {\it "the temperature"}, a so-called {\it "physical temperature"} was proposed when utilizing Tsallis-algebra, see for instance \cite{AbeRef,Refeee1,Refeee2,Refeee3}. It was concluded that if the inverse Lagrange multiplier associated with constrained internal energy is regarded as {\it "the temperature"}, both Tsallis and Clausius entropies become identical. This temperature is believed to differ from the {\it "physical"} one. Based on this assumption, the {\it "physical temperature"} was conjectured to be extrapolated to the real {\it "temperature"}, as it is assumed that this quantity is identical to the freezeout temperature. More details shall be elaborated in section \ref{sec:extraTq}. It was concluded that, the assumption on inverse Lagrange multiplier \cite{AbeRef} follows the same path that nonextensivity. It is only fulfilled through modifications in the thermodynamic intensive and extensive quantities, such as internal energy and temperature. In the present work, we introduce a GNS approach, which assumes that the phase space determines the degree of (non)extensivity depending on two critical exponents defining equivalence classes $(c,d)$, while the intensive and extensive thermodynamic quantities shouldn't necessarily to be modified. To summarize, the ($c, d$)-temperature seems not distinguished from the $T$ parameter of Tsallis statistics. 

In the present paper, generalized thermostatistics (GTT) is applied to investigate the chemical freezout. The main line of the critical remarks on Tsallis type statistics is related to the interpretation of the resulting temperature. The temperature is a well elaborated concept in GTT. Characterizing the statistical nature of the particle production is the main focus of the present work. As introduced, the proposed GNS approach is able to find out either the particle production is to be described by extensive or nonextensive statistics. This basically differs from an {\it ad hoc} implementation of BG and Tsallis approaches. In Ref. \cite{Tawfik:2016pwz}, one of the authors (AT) confronted calculations from the hadron resonance gas (HRG) model with GNS to the corresponding lattice QCD thermodynamics and to various particle ratios measured at $7.7$ and $200~$GeV. He concluded that the lattice QCD simulations are best reproduced when the scaling exponents ($c, d$) get the extensive values, i.e. both are unity, while the particle production seems to have nonextensive classes, i.e. both differ from unity. In other words, the proposed GNS approach seemed to {\it "suggest"} that the lattice calculations are extensive while the particle ratios are stemming from nonextensive processes. Let us first emphasize that this conclusion is fully correct, as the lattice QCD simulations are actually based on extensivity (additivity) of subsystems (quarks and glouns proceeded and communicated by CPUs). On the other hand, the nature of the particle productions in the relativistic heavy-ion collisions might be different. The present work is devoted to a detailed analysis of the particle production in a wide range of beam energies. Accordingly, the statistical approach whether extensive or nonextensive, which well reproduces the chemical freezeout parameters, shall be defined. For the sake of completeness, we highlight that the lattice freezeout parameters are found compatible to the ones deduced from extensive fits of HRG calculations to various particle ratios measured in a wide range of beam energies. This is a solid confirmation that the proposed statistical approach, GNS, works well for the particle productions.

The present paper is organized as follows. The proposed approaches shall be shortly discussed in section \ref{sec:app}. A reminder to GNS shall be given in section \ref{sec:gns}. Sections \ref{sec:cd} and \ref{sec:extraTq} shall be devoted to the characterizations of ($c, d$) nonoextensive-entropy and the extrapolation of $T_q$ to $T_{\mathrm{ch}}$, respectively. GNS in HRG model shall be detailed in section \ref{sec:Generic}. In section \ref{sec:fittingPR}, the statistical fits of various particle-ratios at different beam-energies shall be outlined. The freezeout parameters, the temperatures and the baryon chemical potentials, deduced from the statistical fits of GNS grand-canonical partition function to various particle-ratios measured in heavy-ion experiments shall be determined and compared with the results deduced from extensive HRG calculations.  The resulting equivalence classes $(c, d)$ shall be discussed in \ref{sec:resultingcd}. Section \ref{sec:cncl} outlines the conclusions and the final remarks.

\section{Approaches}
\label{sec:app}

To explore the phase diagram of hot and dense nuclear-matters which can be estimated at the beam-energies  
$3.8$, $4.2$, $5$, $6.4$, $7.7$, $8$, $9$, $11.5$, $12$, $17$, $19.6$, $27$, $39$, $62.4$, $130$, $200$, and $2760~$GeV play an essential role. It was assumed that this can be accomplished from the statistical fit of thermal models, such as HRG, in which GNS approach is implemented \cite{Tawfik:2016pwz,Tawfik:2016jol,Tawfik:2017bul} to various particle-ratios measured in relativistic heavy-ion collisions. 

A detailed descriptions for the generic nonextensive statistics can be found at \cite{Tawfik:2016pwz,Tawfik:2016jol,Tawfik:2017bul}. In the section that follows we give a short reminder to the generic nonextensive statistics.

\subsection{Reminder to generic nonextensive statistics}
\label{sec:gns}

Shannon entropy ($S_1$) is uniquely characterized by four axioms, Khinchin axioms:
\begin{enumerate}
\item SK1, continuity: for any $n\in  \mathbb{N}$, $S_1(p)$ is continuous with respect to $p\in\Delta_n$, i.e. the entropy depending on $p$, contentiously. Accordingly, function $g(p)$ becomes continuous, as well.
\item SK2, maximality: for given $n\in  \mathbb{N}$ and for $(p_1,\cdots,p_n)\in\Delta_n$, $S_1(p_1,\cdots,p_n)$ reaches maximum at $p_i=1/n$, where $i=1,\cdots,n$. Consequently, for equi-distribution ($p$), the entropy becomes maximum and the function $g(p)$ become a concave function.
\item SK3, expandability: $S_1(p_{1},\cdots,p_{n},0) = S_1(p_{1},\cdots,p_{n})$, i.e. adding a non-zero probability state does not change the entropy so that $g(0)=0$.
\item SK4, generalized additivity: if $p_{ij} \geq 0$, $p_i=\sum_{j=1}^{m_i} p_{ij}$, and $\sum_{i=1}^{n} p_{i}=1$, where $i=1,\cdots,n$ and $j=1,\cdots,m_i$, then $S_1(p_{11},\cdots,p_{nm_n}) = S_1(p_{1},\cdots,p_{n}) + \sum_{i=1}^{n} p_i S_1\left(\frac{p_{i1}}{p_i}\cdots,\frac{p_{i m_i}}{p_i}\right)$. This means that the entropy of a system, which is divided into two subsystems $A$ and $B$, is given as $S_A$ plus the expectation value of $S_B$ conditional on $A$.
\end{enumerate}
Fulfilling the four axioms characterizes ($c, d$) extensive entropy, where the equivalence classes ($1, 1$) recall BG statistics. Violating the fourth axiom, individually, characterizes ($c, d$) nonextensive entropy. The latter can be Tsallis, where ($q, 0$), or GNS, where ($c, d$) take other values. We recall again that, the scaling exponents $c$ and $d$ are corresponding to two  independent asymptotic properties of the entropy functional ($s$), namely \cite{Thurner1,Thurner2}
\bea
z^c &=& \lim_{x\rightarrow 0} \frac{s(z x)}{s(x)}, \\
(1+a)^d &=& \lim_{x\rightarrow 0} \frac{s(x^{1+a})}{x^{a c}s(x)},
\eea
with $a$ and $z$ are arbitrary variables ranging between $0$ and $1$, but not affecting the (non)extensivity.

\subsection{Characterizations of ($c$,$d$) nonextensive entropy}
\label{sec:cd}

From {\it formed} processes, on which the phase space depends, the extensive entropy can be computed, where the phase space linearly increases with the increase in the system size ($N$). On the other hand, for {\it deformed} processes, one needs a general  representation of the entropy functional ($s$). Generalized families of class-representation can be achieved through two-time differentiable, monotonically increasing functions. For instance, a generalization for Shannon entropy was proposed  as $H(p)=\sum_{i=1}^{W} h(p_i)$, where $W$ gives the number of states of the phase-space processes \cite{Thurner1}. From the asymptotic properties of the trace-form entropy functionals, following representations have been proposed \cite{Thurner1}
\begin{enumerate}
\item $G$-representation: $G(p)=\sum_{i=1}^{W} g(p_i)$, where
\bea
g(p_i) &\equiv & - p_i \Lambda(p_i), \label{eq:genG}
\eea
and $\Lambda$ is a generalized logarithm, see Eq. (\ref{eq:genLog}).
\item  $S$-representation: From $-\int_0^{p_i} dy \log(y)=-p_i  \log(p_i) + p_i$, the second term in right-hand side can be absorbed in $-\alpha(\sum_i p_1-1)$, when the normalization constrains of the maximum-entropy-principle is implemented. $\alpha$ is a Lagrangian multiplier. Accordingly, $h(p_i)=-p_i  \log(p_i)$ and
\bea
s(p_i) &=& -\int_0^{p_i} dx\,  \Lambda(x). \label{eq:genH}
\eea
\end{enumerate}
Both representations should give equivalent ($c$,$d$) entropy classes
\bea
s_{(c,d,r)}(x) &=& \frac{r}{c} A^{-d} \exp(A) \Gamma(d+1,A-c \log(x)) - r x,
\eea
where $A=c d r /(1-r(1-c))$. Correspondingly, both exponential and logarithm should be generalized. The earlier shall be introduced in Eq. (\ref{eq:epsln}), while the latter reads
\bea
\Lambda_{(c,d,r)}(x) &=& r \left[1-x^{c-1}\left(1- \frac{1-r(1-c)}{d r } \log(x)\right)^d\right]. \label{eq:genLog}
\eea
 
Now we are left with the fifth property of ($c$,$d$) nonextensive entropy. This should be characterized by vanishing entropy property,
\bea
0 &=& \int_{0}^{b_{(c,d,r)}} dx \Lambda_{(c,d,r)}(x), \label{eq:ve1}\\
S_{(c,d,r)}(p) &=& \sum_{i=1}^{W} s_{(c,d,r)}(b_{(c,d,r)} p_i), \label{eq:ve2}
\eea
where $b_{(c,d,r)}$ optimizes the asymptotic extensivity. When scaling $s_{(c,d,r)}$, even the representation given in Eq. (\ref{eq:genG}) fulfils vanishing-entropy property, as well.

To summarize, besides SK1, SK2, and SK3, the ($c$,$d$) nonextensive entropy is characterized by:
\begin{itemize}
\item trace-form of the entropy functional, $S(p)=\sum_{i=1}^W s(p_i)$, and
\item vanishing entropy property, Eqs. (\ref{eq:ve1})-(\ref{eq:ve2}).
\end{itemize}

\subsection{Extrapolation of $T_q$ to $T_{\mathrm{ch}}$}
\label{sec:extraTq}

An {\it ab initio} assumption was made that the Tsallis {\it freezeout} temperature can be extrapolated to the  BG-temperature ($T_{\mathrm{ch}}$) deduced from extensive thermal-statistical models \cite{deppmn0,deppmn1},
\bea
T_q &=& T_{\mathrm{ch}} + (q-1) k, \label{eq:Tq1}
\eea
where $q$ is Tsallis nonextensive parameter, i.e. $q>1$. $k$ is conjectured to depend on the energy transfer between a source and the surrounding. It was found that at $T_{\mathrm{ch}} =192\pm 15~$MeV, $k=-(950\pm10)~$MeV \cite{deppmn2}. As mentioned in section \ref{sec:intro}, The temperature deduced from the statistical fits of the transverse momentum distributions was interpreted as the kinetic freezeout temperature. To remain within the scope of the present paper, we leave this point without any further discussion.

In a classical ideal gas, the Tsallis entropy, which describes an exact entropy for microcanonical distribution of such  gas, was proposed to relate Tsallis temperature, the $q$-temperature, ($T_q$) to the BG freezeout temperature ($T_{\mathrm{ch}}$) \cite{tamas1}
\bea
T_q &=& T_{\mathrm{ch}} \, \exp\left(\frac{S_q}{C}\right), \label{eq:Tq2}
\eea
where $C$ is the heat capacity of reservoir system and $S_q$ is Tsallis-entropy.

To summarize, the $q$-temperature, which as discussed in section \ref{sec:intro} is know as {\it "physical temperature"}, is about $2-3$ times smaller than $T_{\mathrm{ch}}$ know as {\it "the temperature"} \cite{Deppmann2015}. Accordingly, the resulting freezeout phase-diagram largely differs from the one drawn from the extensive statistical fits of various measured particles-ratios and recent lattice QCD simulations, especially the freezeout temperature \cite{Deppmann2015}. Despite the well-know sign-problem in the lattice QCD calculations at finite chemical potential ($\mu_b$), where the Monte Carlo techniques likely fails, both deconfinement and freezeout temperatures are conjectured to be compatible with each other, especially at small chemical potentials \cite{Ding:2015ona}. This range of $\mu_b$ is to be related to RHIC and LHC energies, where precise measurements for different particle-ratios are available \cite{Tawfik:2013bza,Tawfik:2014dha}. What is obtained so-far is that at vanishing $\mu_b$ and at the corresponding $T_q$, the resulting HRG thermodynamics, even when applying Tsallis-algebra, considerably differs from the one based on the first-principle lattice calculations. This can be understood from the fact that the latter assume extensivity and additivity besides an overall thermal equilibrium. In light of this, it is apparent that the Tsallis-type nonextensivity shouldn't be utilized in order to reproduce of the lattice calculations. In this regards, we first remind with the remarkable success of HRG with BG statistics in reproducing the first-principle lattice thermodynamics \cite{Karsch:2003vd,Karsch:2003zq,Redlich:2004gp,Tawfik:2004vv,Tawfik:2004sw,Tawfik:2005gk,Tawfik:2005qh}. This is another argumentation of why the proposed GNS approach is the proper one. It reproduces the lattice calculations only when the equivalence classes ($c, d$) get extensive values, i.e. ($1,1$). The success in characterizing the statistical nature of the lattice calculations manifests the fact that the proposed approach (GNS) is generic and its applicability is much wider than that of BG and Tsallis.

As discussed, comprehensive works have been conducted in order to bring {\it "physical temperature"} closer to the {\it "the temperature"}, for instance Eqs. (\ref{eq:Tq1})-(\ref{eq:Tq2}). These works assume that the nonextensivity in the particle production can be characterized through a radical change in intensive or extensive thermodynamic quantities, such as the internal energy and the temperature. The resulting $T_q$, which is lower than $T_{\mathrm{ch}}$, are extrapolated. One of authors (AT) has first introduced GNS approach to the statistical fits of particle ratios at $200$ and $2760~$GeV \cite{Tawfik:2016pwz}. The present work covers a wider range of beam energies. Accordingly, the two critical exponents defining the equivalence classes $(c, d)$ can be determined. While both intensive or extensive thermodynamic quantities remain almost unchanged, $(c, d)$ unambiguously characterize the nonextensive nature of the particle production. In a future work, we shall study the energy dependence of $(c, d)$.

\subsection{GNS in HRG model}
\label{sec:Generic}

As introduced in Ref. \cite{Tawfik:2016pwz,Tawfik:2016jol,Tawfik:2017bul}, both extensive and nonextensive statistical properties can be determined by two scaling exponents defining equivalence classes $(c, d)$ for both correlated and uncorrelated systems, for instance. In its thermodynamic limit, the statistical nature of the system of interest can be characterized, unambiguously. The remarkable success of the thermal models (extensive) \cite{Tawfik:2014eba,Tawfik:2010aq} in describing the particle production and that of the Tsallis nonextensive algebra \cite{ReFF1,ReFF2} simply mean that the cost which should be paid, namely the radical reduction in the freezeout temperature, was not necessarily. Such an {\it ad hoc} implementation of statistical approaches is not the best way to decide whether the particle ratios or transverse momentum distributions or any other phenomenological spectra are extensive or nonextensive.  A direct confrontation with GNS is this \cite{Tawfik:2015}. Here, the system is not enforced to be biased towards either extensive or  nonextensive statistical description. With a generalized exponential function, Eq. (\ref{eq:epsln}),  
\bea
\ln\, Z(T) &=& \pm V\, \sum_i^{N_{\mathtt{M|B}}}\, \frac{g_i}{(2\, \pi)^3}\, \int_0^{\infty}\, \ln\left[1\pm\varepsilon_{c,d,r}(x_i)\right]\; d^3\, {\bf p}, \label{eq1}
\eea
where $V$ is the fireball volume and $x_i=[\mu_{i}-E_i({\bf p})]/T$ with $E_i(p)=\sqrt{{\bf p}^2+m_i^2}$ being the dispersion relation of $i$-th state (particle) and $\pm$ represent fermions and bosons, respectively.
The exponential function $\varepsilon_{c,d,r}(x_i)$ is generalized  as \cite{Thurner1,Thurner2}
\bea
\varepsilon_{(c,d,r)}(x) &=& \exp\left\{-\frac{d}{c-a}\left[W_k\left(B\left(1-x/r\right)^{1/d}\right)-W_k(B)\right]\right\}, \label{eq:epsln}
%\varepsilon_{c,d,r}(x_i)=\exp\left[ \frac{-d}{1-c} \left(W_k\left[B\left(1-\frac{x_i}{r}\right)^{\frac{1}{d}}\right]-W_k[B]\right)\right], \label{eq:epsln}
\eea
where $W_k$ is Lambert $W$-function which has real solutions at $k=0$ with $d\geq 0$ and at $k=1$ with $d<0$, and
\be
B=\frac{(1-c)r}{1-(1-c)r} \exp\left[\frac{(1-c)r}{1-(1-c)r}\right],
\ee
where $c$ and $d$ are two critical exponents defining equivalence classes for all interacting and noninteracting systems. They give estimations for two scaling functions with two asymptotic properties. $r$ is almost a free parameter. It is assumed not affecting the ($c$, $d$)-class. %$r=(1-c+c\,d)^{-1}$
\begin{itemize}
\item for $d>0$, $r<1/(1-c)$,
\item for $d=0$, $r=1/(1-c)$, and
\item for $d<0$, $r>1/(1-c)$.
\end{itemize}
Alternatively, a particular function for $r$, namely $r=(1-c+c\,d)^{-1}$ was proposed \cite{Thurner1,Thurner2}.

From the partition function, Eq. \eqref{eq1}, the thermodynamical properties such as pressure ($p$) and number density ($n$) \cite{RAFELSKI} can be derived,
\bea
\textit{p} &=& \sum_{i=1}^{N_{\mathtt{M|B}}}\, \frac{g_i\, T}{2 \pi^2} \int_0^\infty  \ln\left[1\pm\varepsilon_{c,d,r}(x_i)\right]\; \textbf{p}^2 d\textbf{p}, \label{FDp}\\
\textit{n} &=& \pm \sum_{i=1}^{N_{\mathtt{M|B}}}\, \frac{g_i}{2 \pi^2} \int_0^\infty \frac{\varepsilon_{c,d,r}(x_i) \; W_0\left[B(1-\frac{x_i}{r})^{\frac{1}{d}}\right]}{(1-c)\left[1\pm\varepsilon_{c,d,r}(x_i)\right] \left(r-x_i\right) \left(1+W_0\left[B(1-\frac{x_i}{r})^{\frac{1}{d}}\right]\right)}\; \textbf{p}^2 d\textbf{p}. \label{FDn1}
\eea
It should be remarked that in Eq. (\ref{eq1}), the logarithmic function should also be generalized. This was given in Eq. (\ref{eq:genLog}), when assuming $G$-representation.

In the present calculations, the possible decay channels of heavy resonances are taken into consideration. The $k$-th particle final number density  (identical expressions for other thermodynamic quantities, such as pressure, entropy and energy density, can be deduced) is given as
\bea
n_k^{final} &=&n_k +\sum_{l\neq k}^{N_{\mathtt{M|B}}}\; b_{l \rightarrow k}\, n_l, \label{eq:nn1}
\eea
where $b_{l \rightarrow k}$ is the effective branching ratio of $l$-th hadron resonance into the $k$-th particle of interest. It is noteworthy mentioning that the resonance decays might be seen as correlations and  interactions among the produced particles and thus the introduction of nonextensivity to the particle production in final state becomes eligible.

\section{Results}
\label{sec:reslt}

\subsection{Chemical freezeout phase-diagram}
\label{sec:fittingPR}

The particle ratios measured by the E866, NA44, NA49, and NA57 experiments at the Superproton Synchrotron (SPS)  \cite{Ahle:2000wq,Klay:2001tf,Ahle:1999uy,Andronic:2005yp} at %$3.2$, 
$3.8$, $4.2$, $5$, $6.4$, $8$, $9$, $12$, and $17~$GeV and by the STAR experiment at the Relativistic Heavy Ion Collider (RHIC) at $7.7$, $11.5$, $19.6$, $27$, $39$, $130$, and $200~$GeV shall be compared with the HRG calculations. Together with the ALICE measurements, we construct a data set covering beam energies from $3.8$ up to $2760~$GeV.

\begin{figure}[hbt]
\includegraphics[width=8.cm]{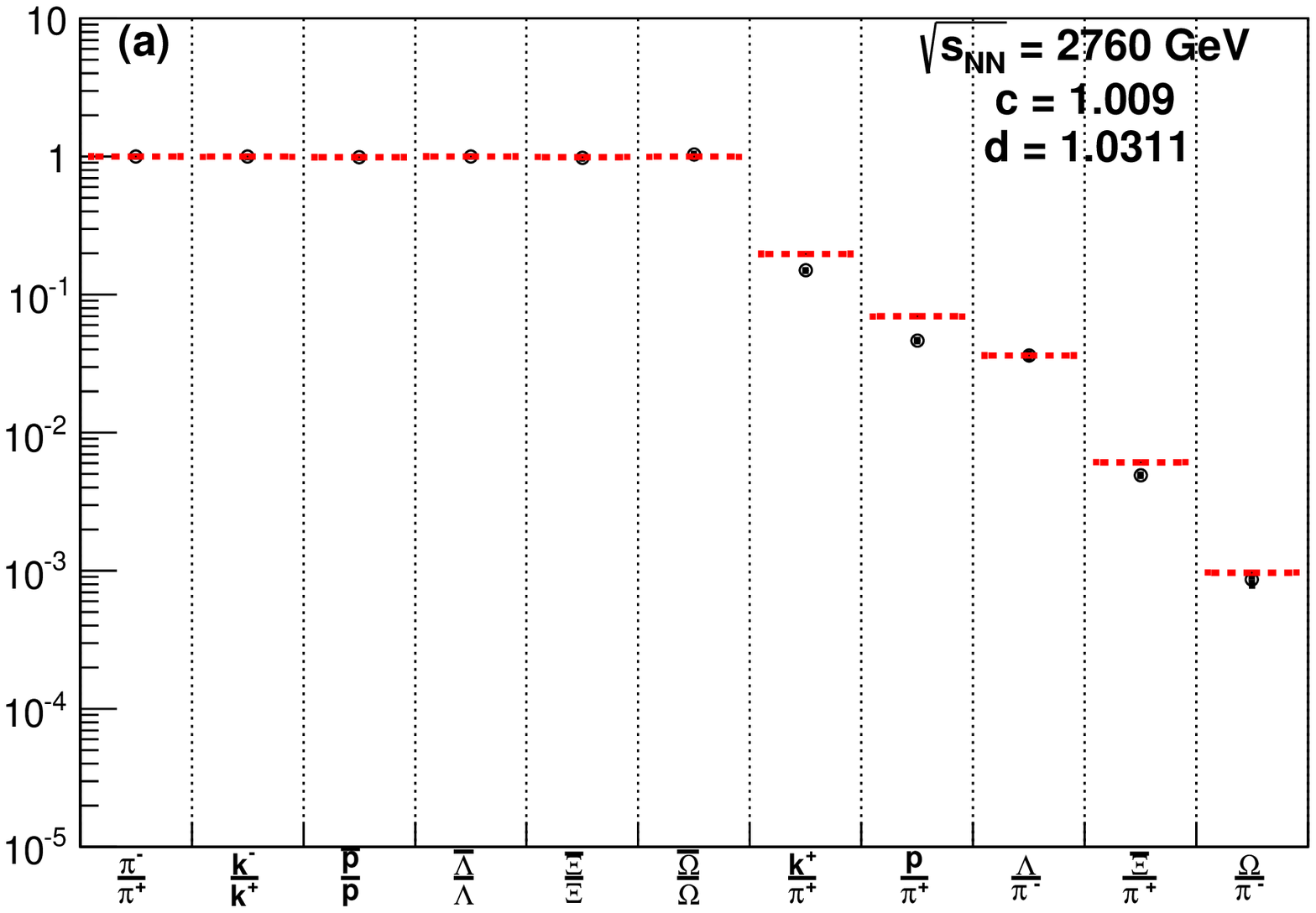}
\includegraphics[width=8.cm]{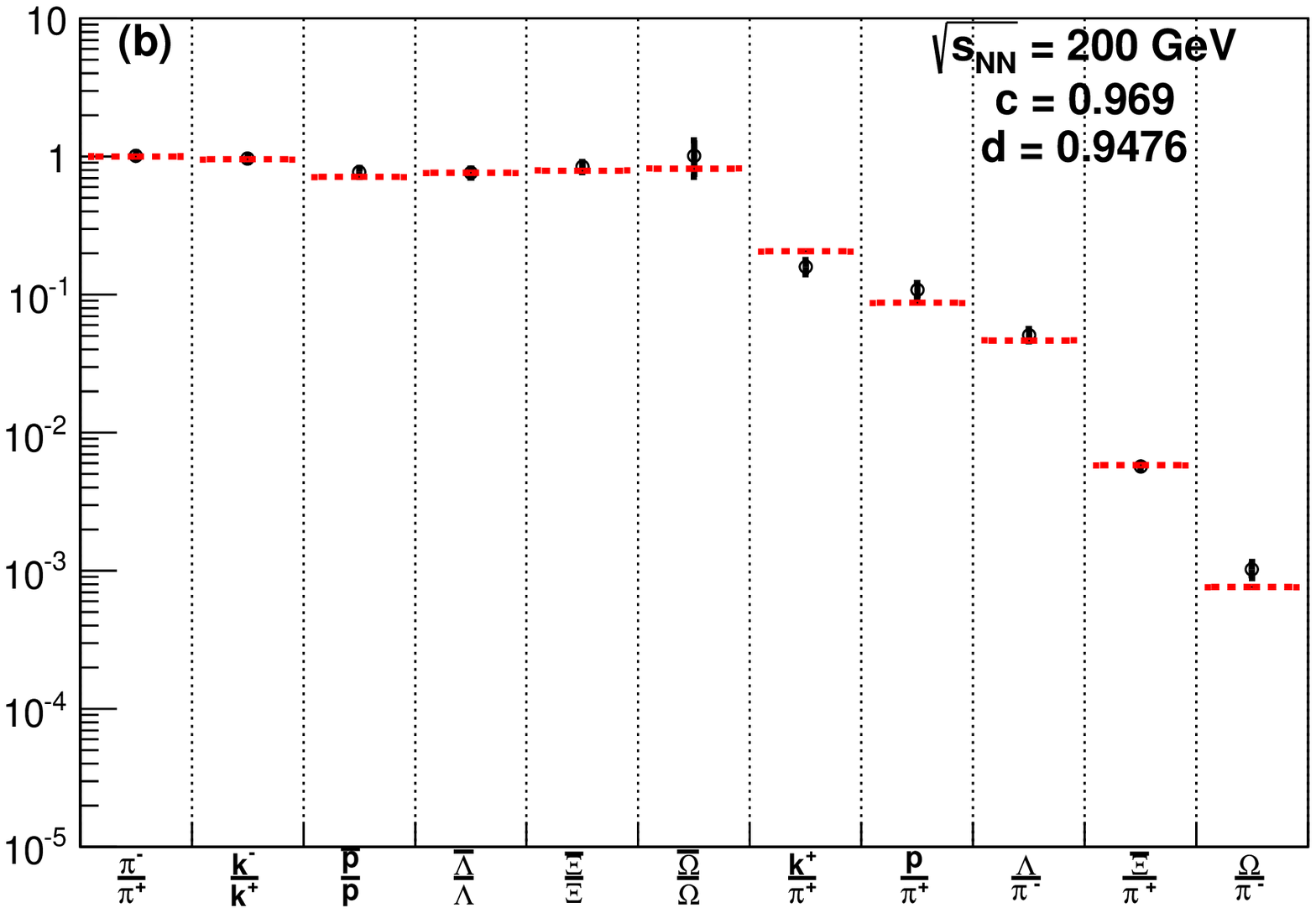}
\caption{Left panel (a): different particle ratios measured at $2760~$GeV (symbols) are fitted to HRG with GNS (dashed lines). Right panel (b) shows the same but at $200~$GeV. The resulting exponents ($c$, $d$) and $\chi^2$ are given in top right corners and Tab. \ref{Tab1}. The freezeout temperature and chemical potential are also listed in Tab. \ref{Tab1}. \label{Fig1}}
\end{figure}

\begin{figure}[h!]
\includegraphics[width=7.cm]{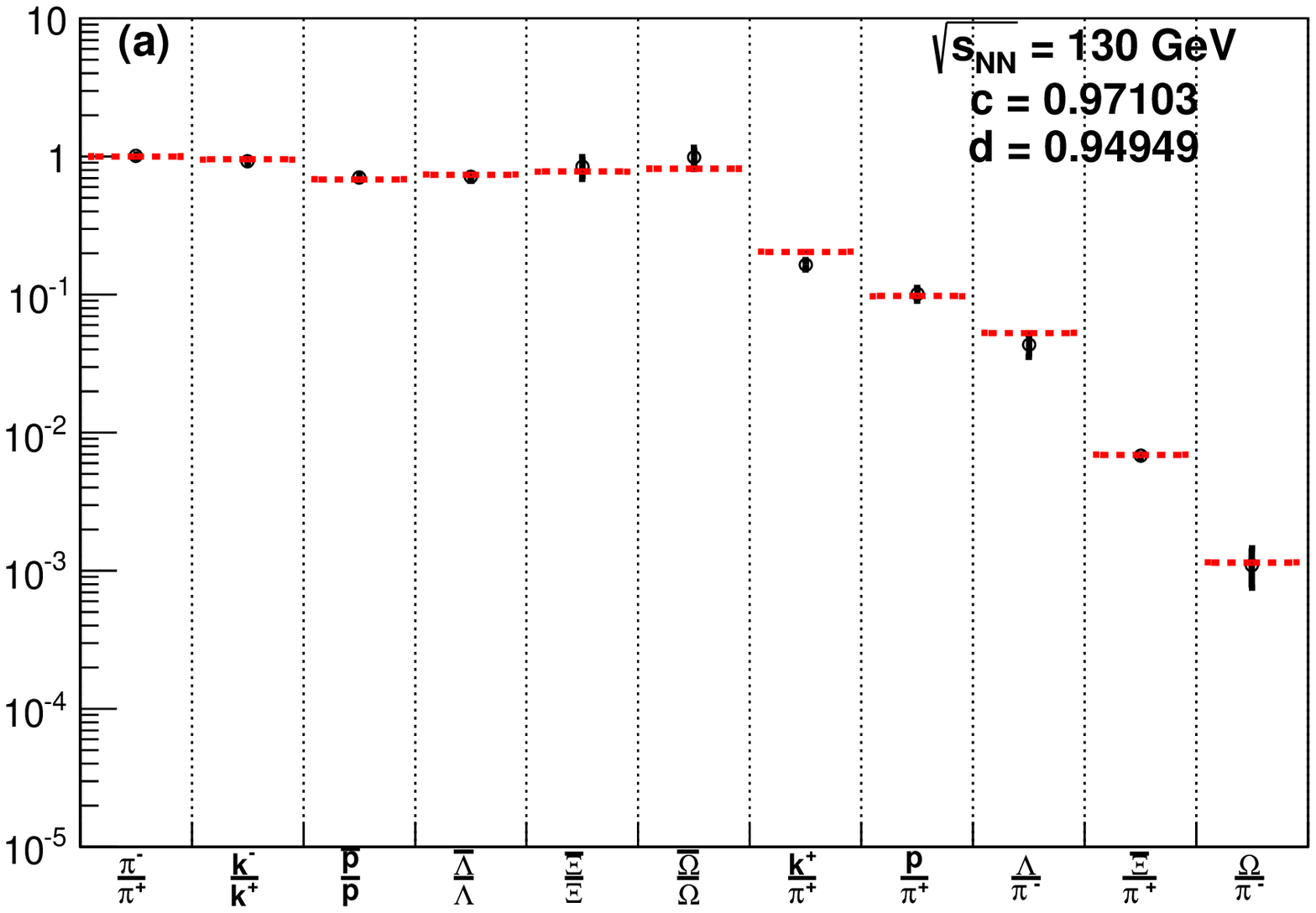}
\includegraphics[width=7.cm]{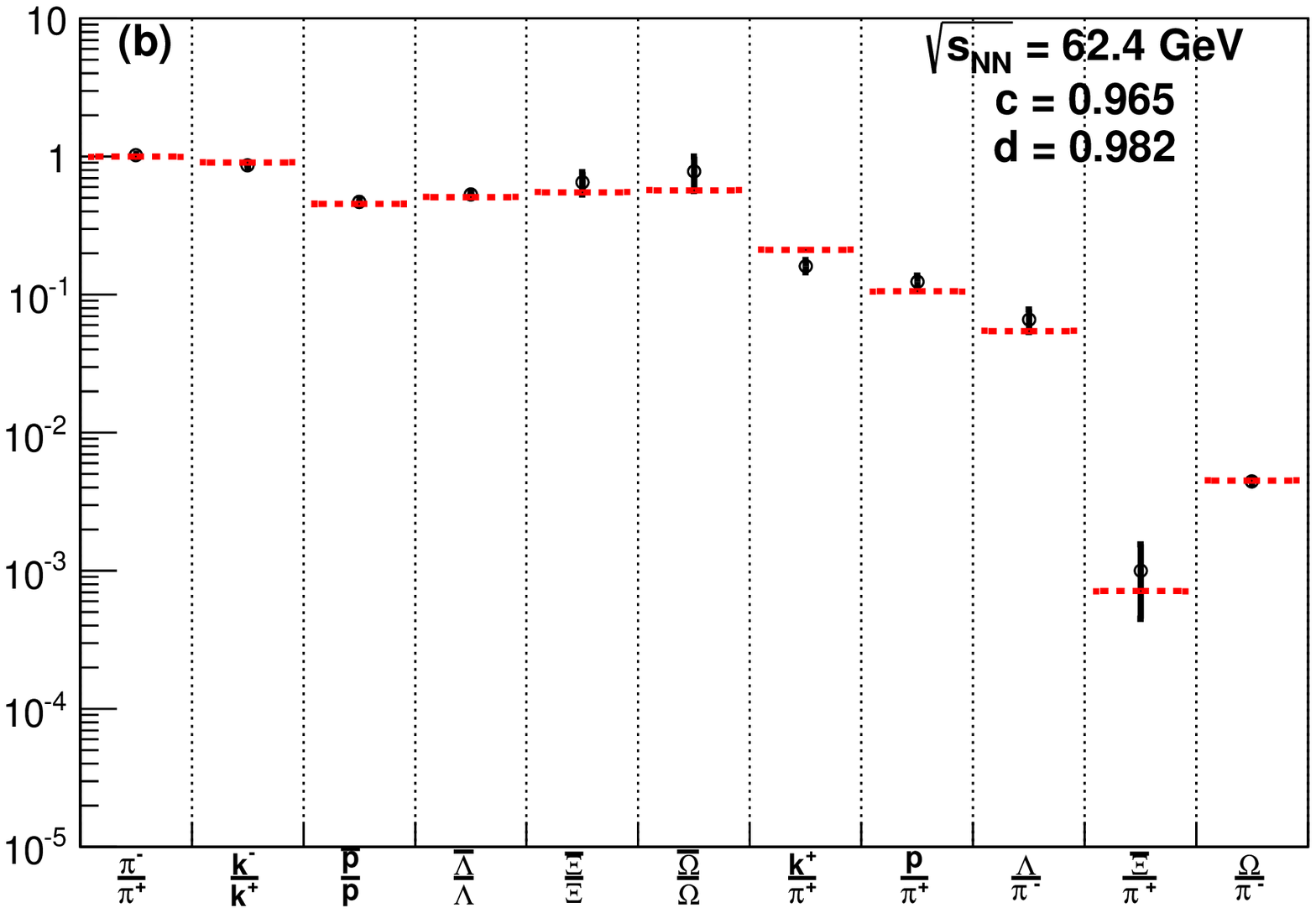}\\
\includegraphics[width=7.cm]{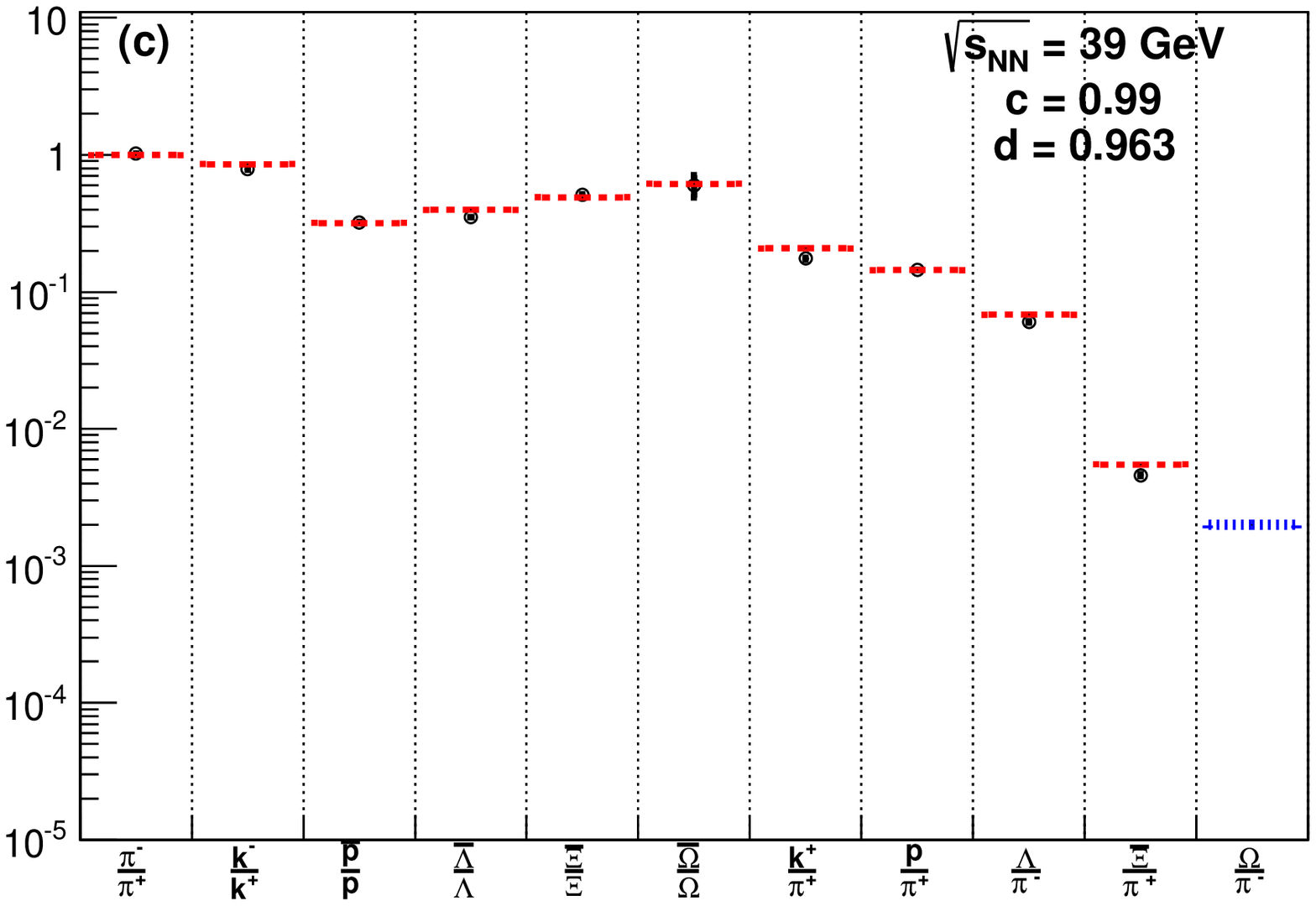}
\includegraphics[width=7.cm]{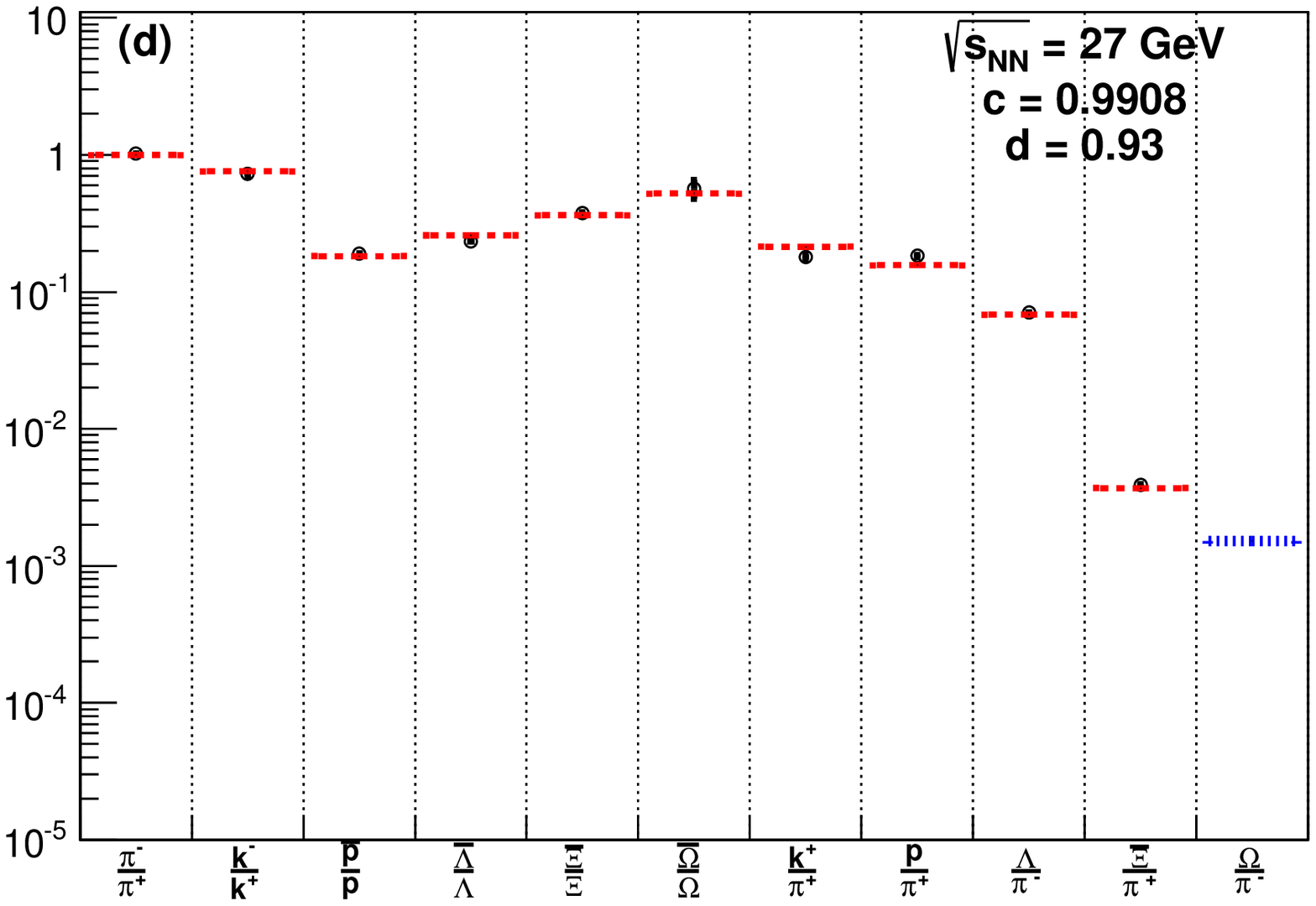}\\
\includegraphics[width=7.cm]{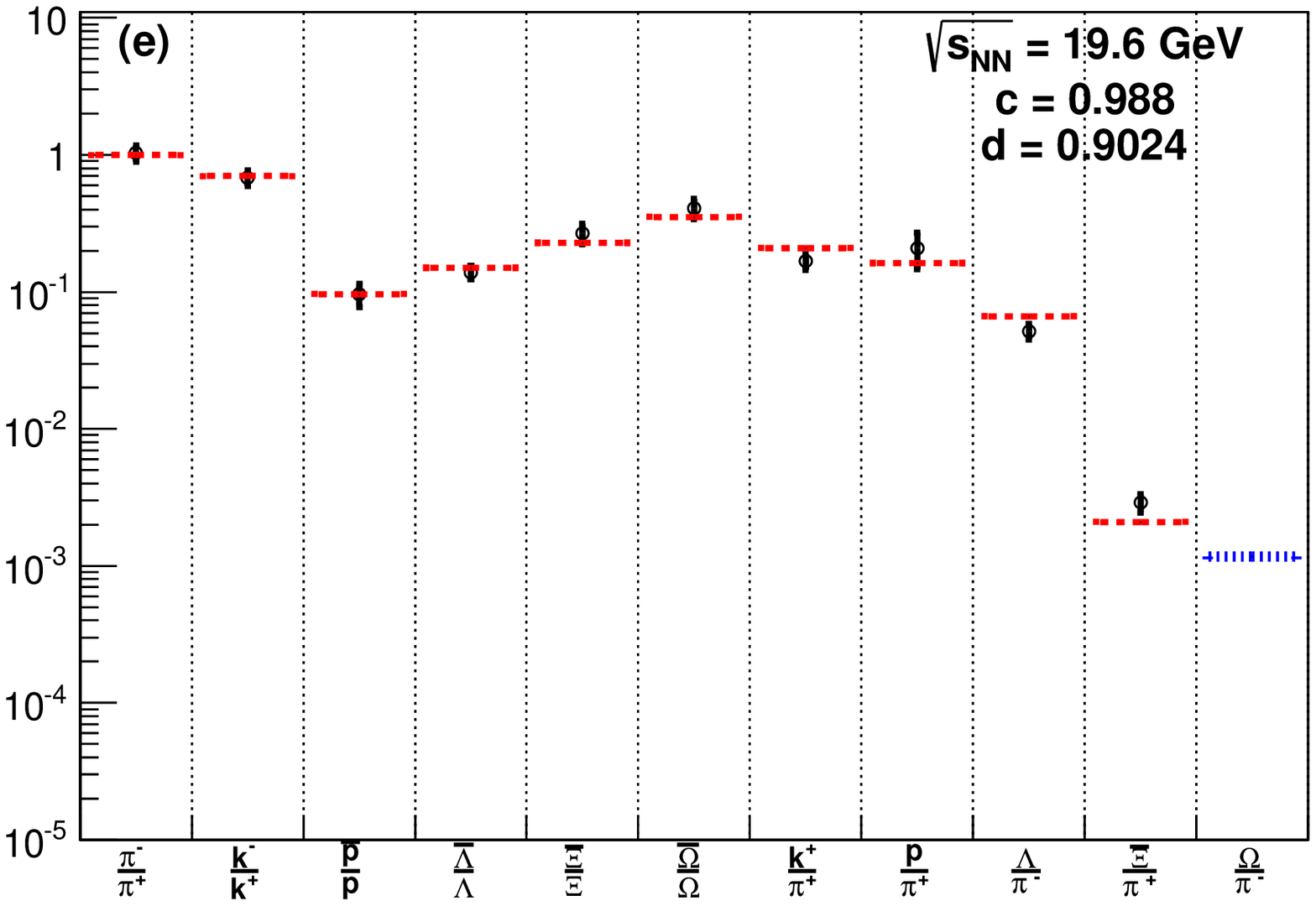}
\includegraphics[width=7.cm]{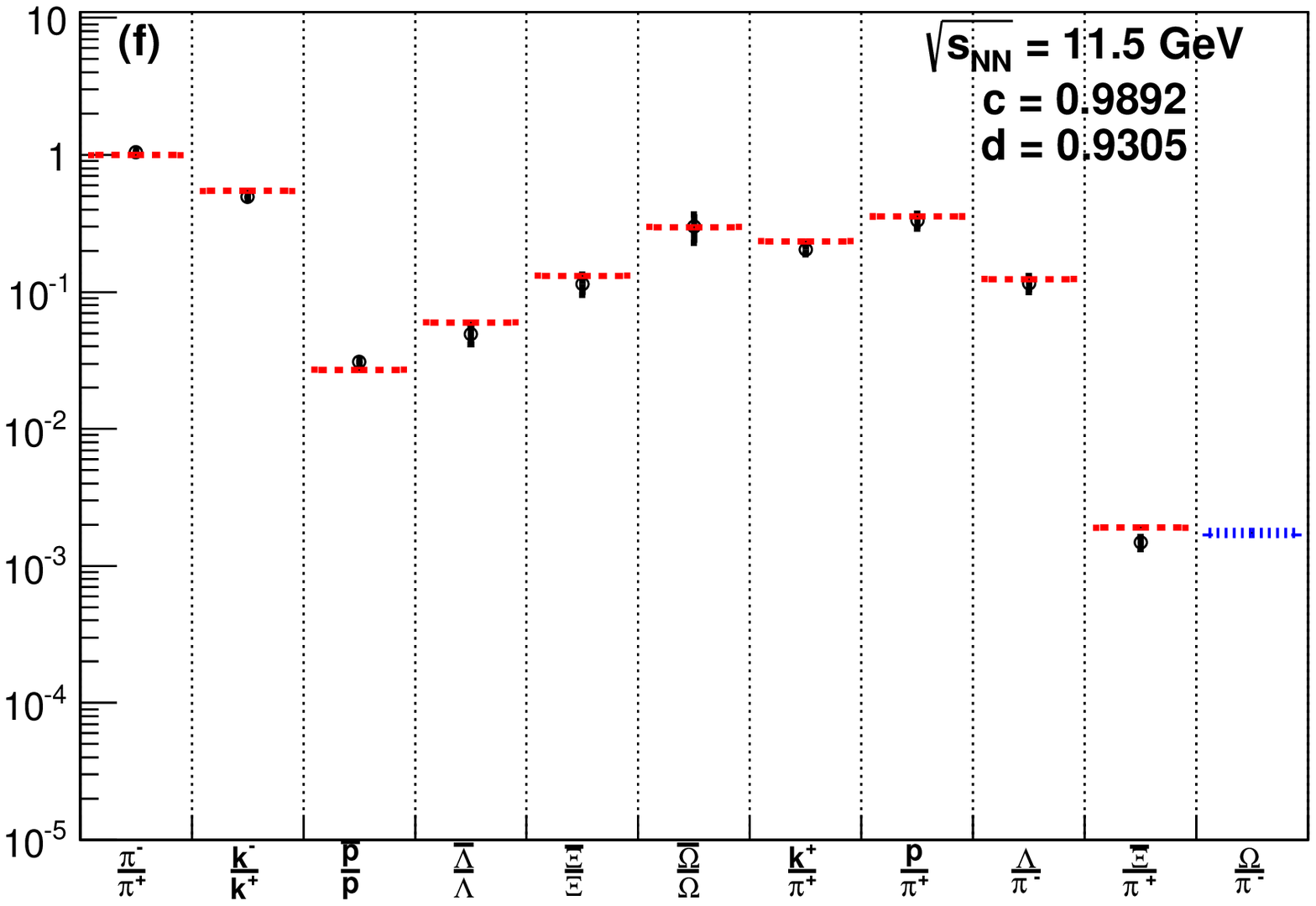}\\
\includegraphics[width=7.cm]{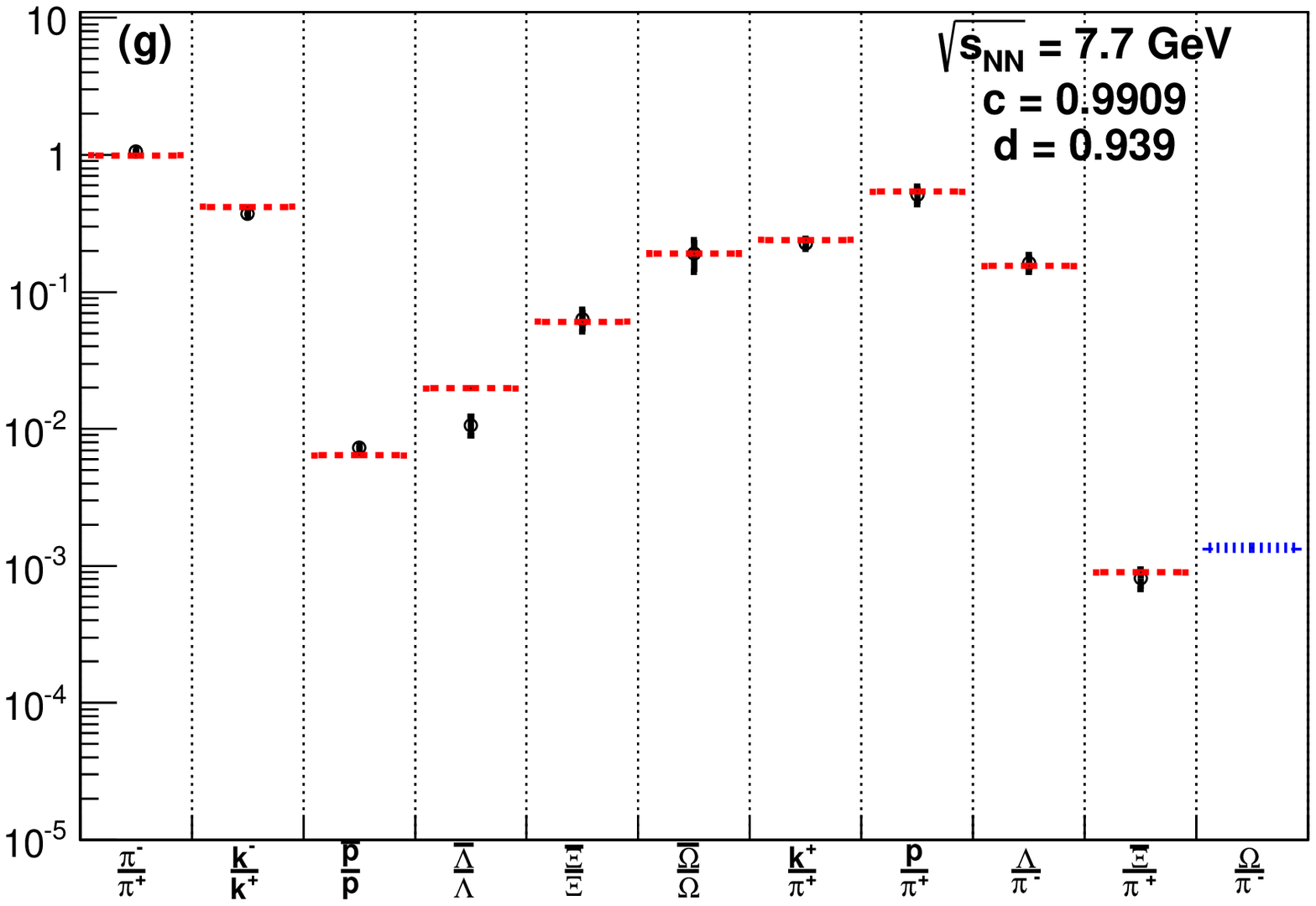}
\caption{Panel (a): different particle-ratios deduced from GNS fits (dashed lines)  to the experimental results at $130~$GeV (symbols). Panels (b), (c), (d), (e), (f), and (g) show the same but at $62.4$, $39$, $27$, $19.6$, $11.5$, and $7.7~$GeV, respectively. The scaling exponents ($c$, $d$) and $\chi^2$ are given in top right corners and Tab. \ref{Tab1}. \label{RHIC}}
\end{figure}

\begin{figure}[h!]
\includegraphics[width=7.cm]{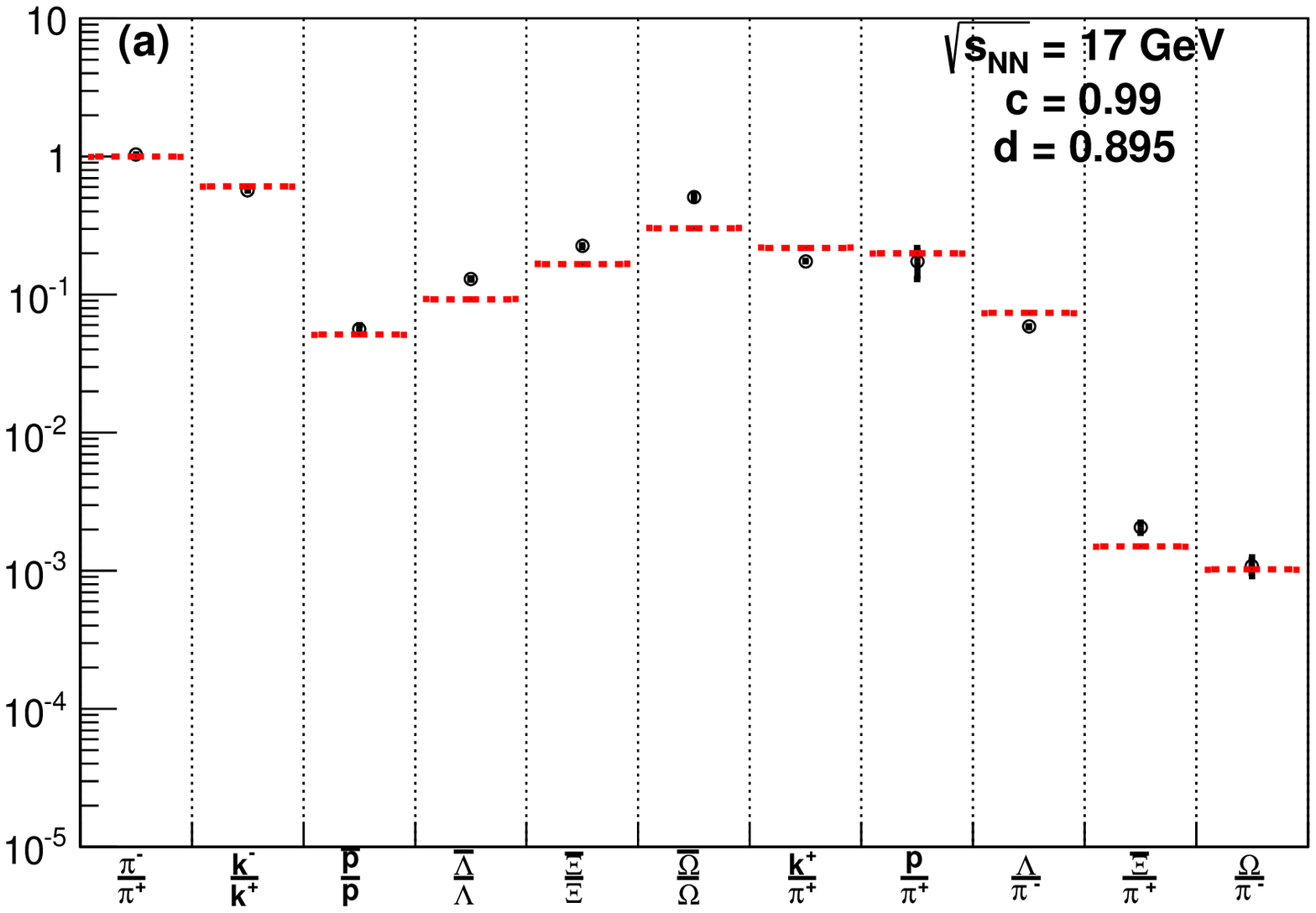}
\includegraphics[width=7.cm]{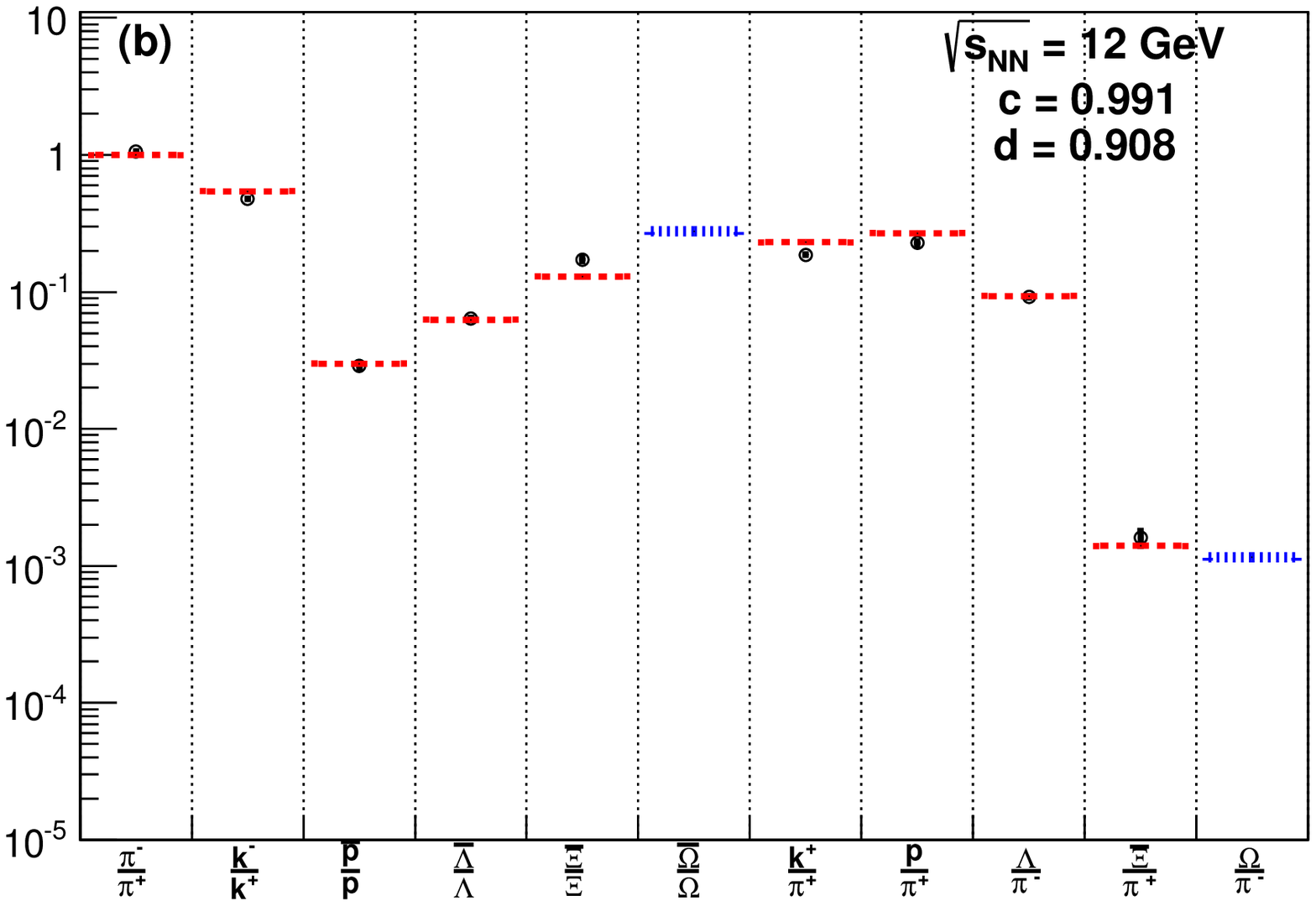}\\
\includegraphics[width=7.cm]{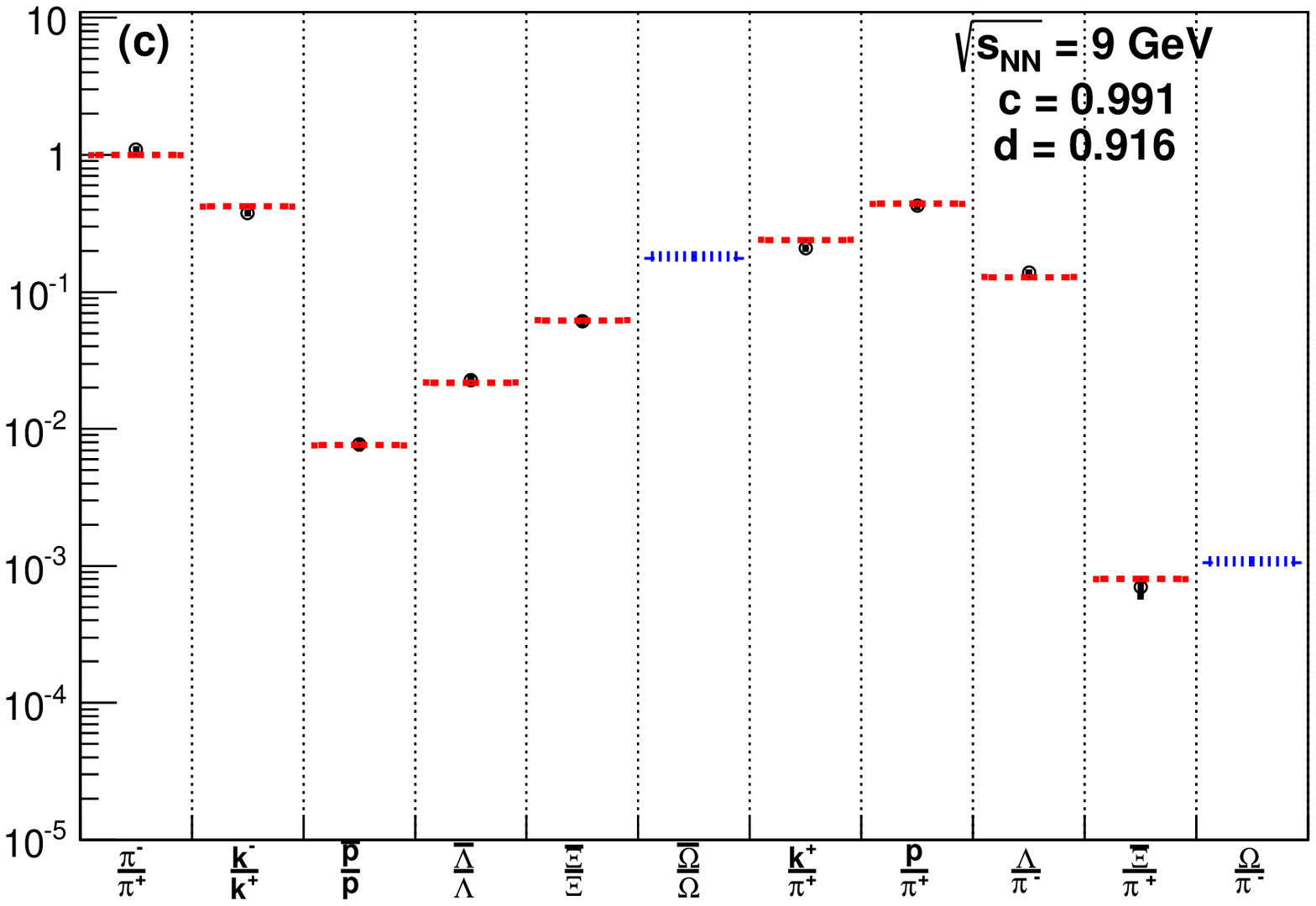}
\includegraphics[width=7.cm]{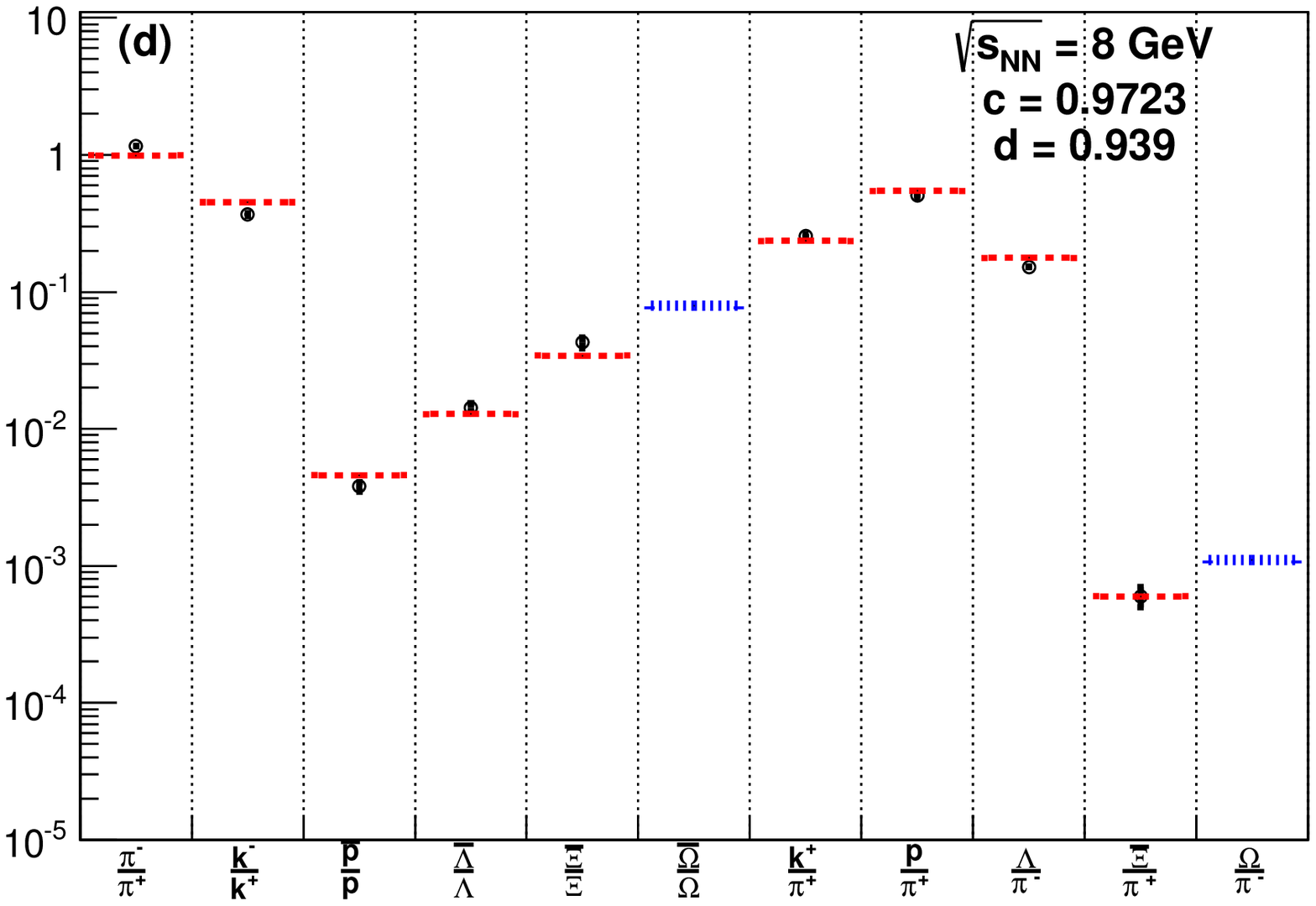}\\
\includegraphics[width=7.cm]{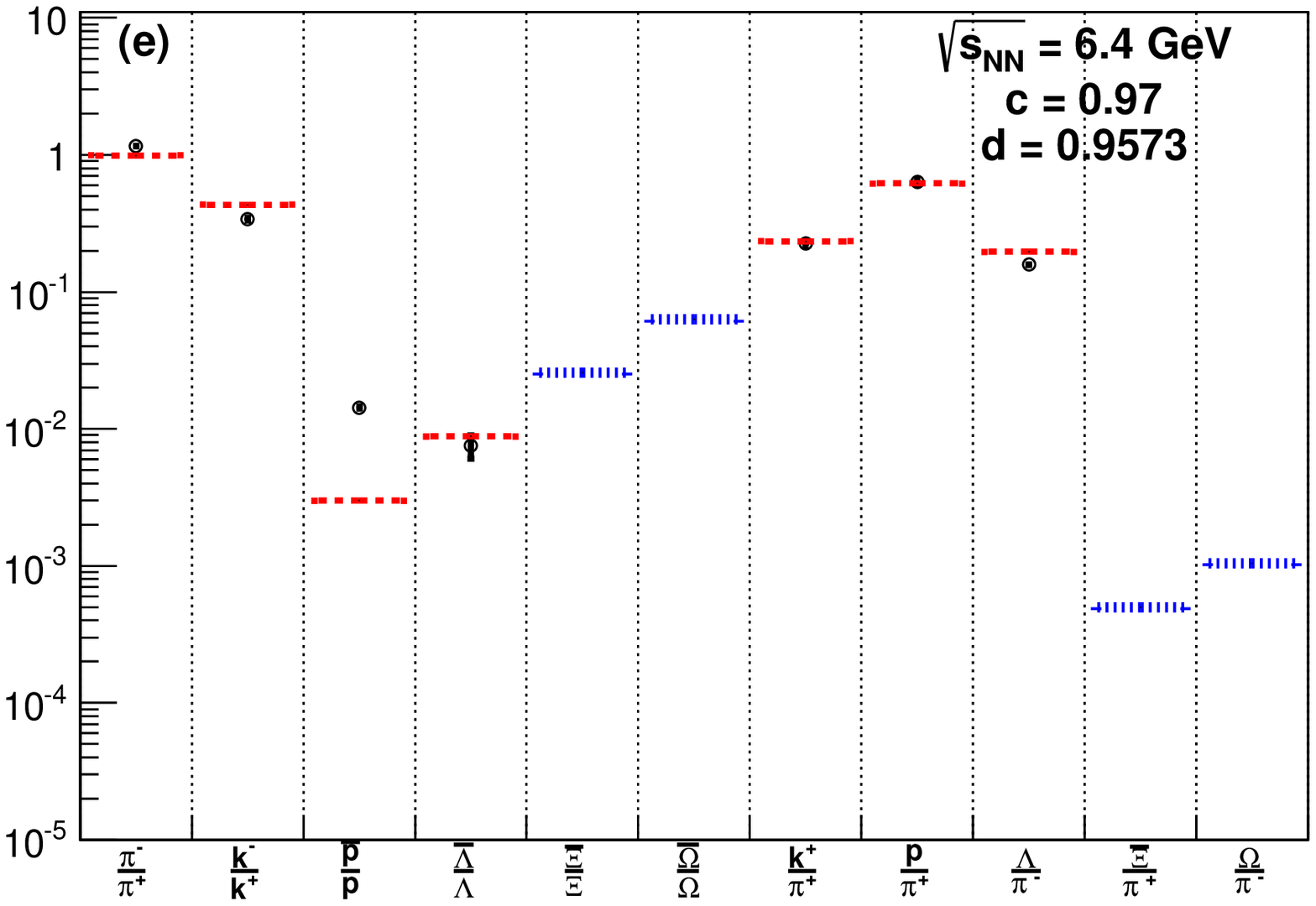}
\includegraphics[width=7.cm]{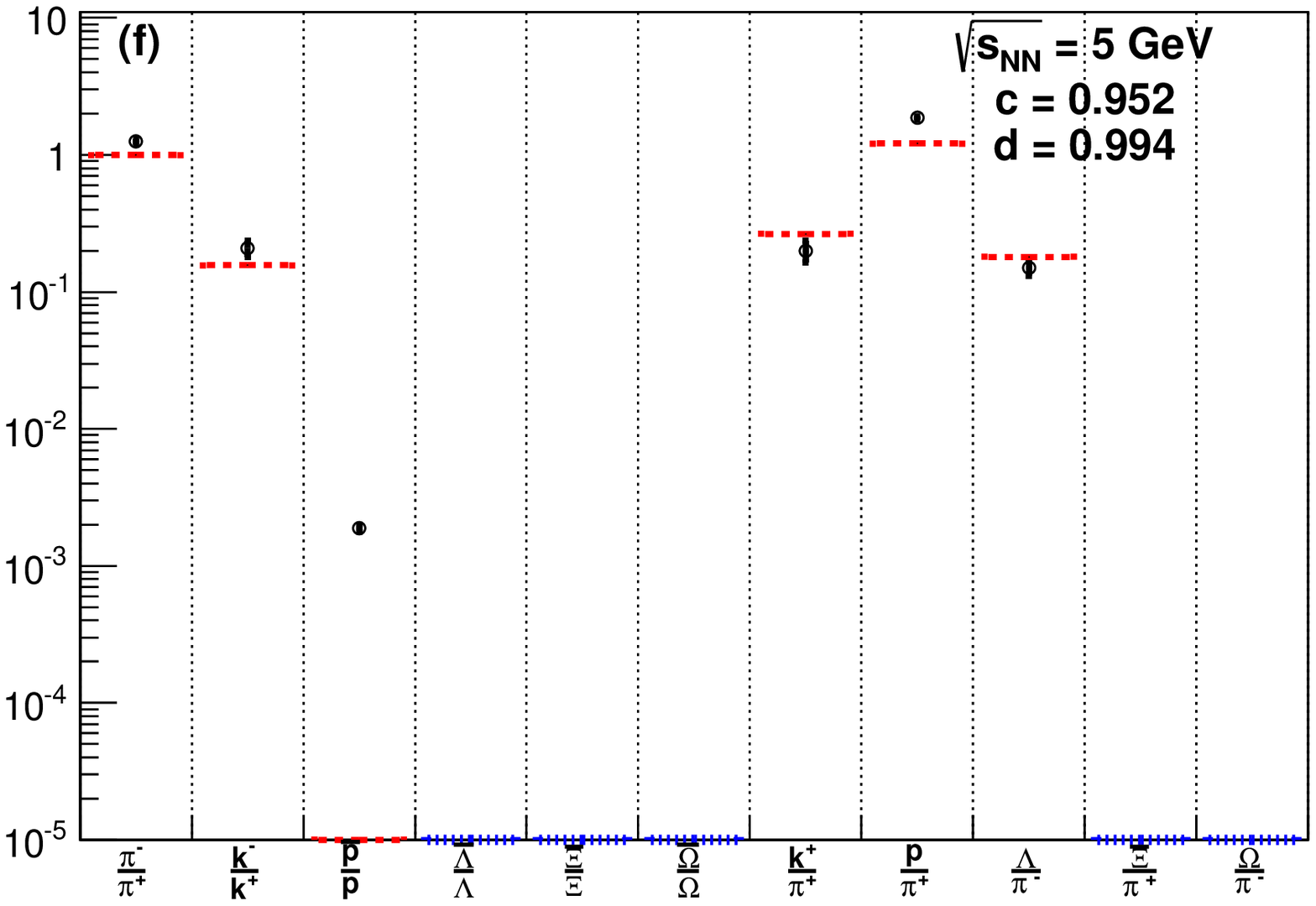}\\
\includegraphics[width=7.cm]{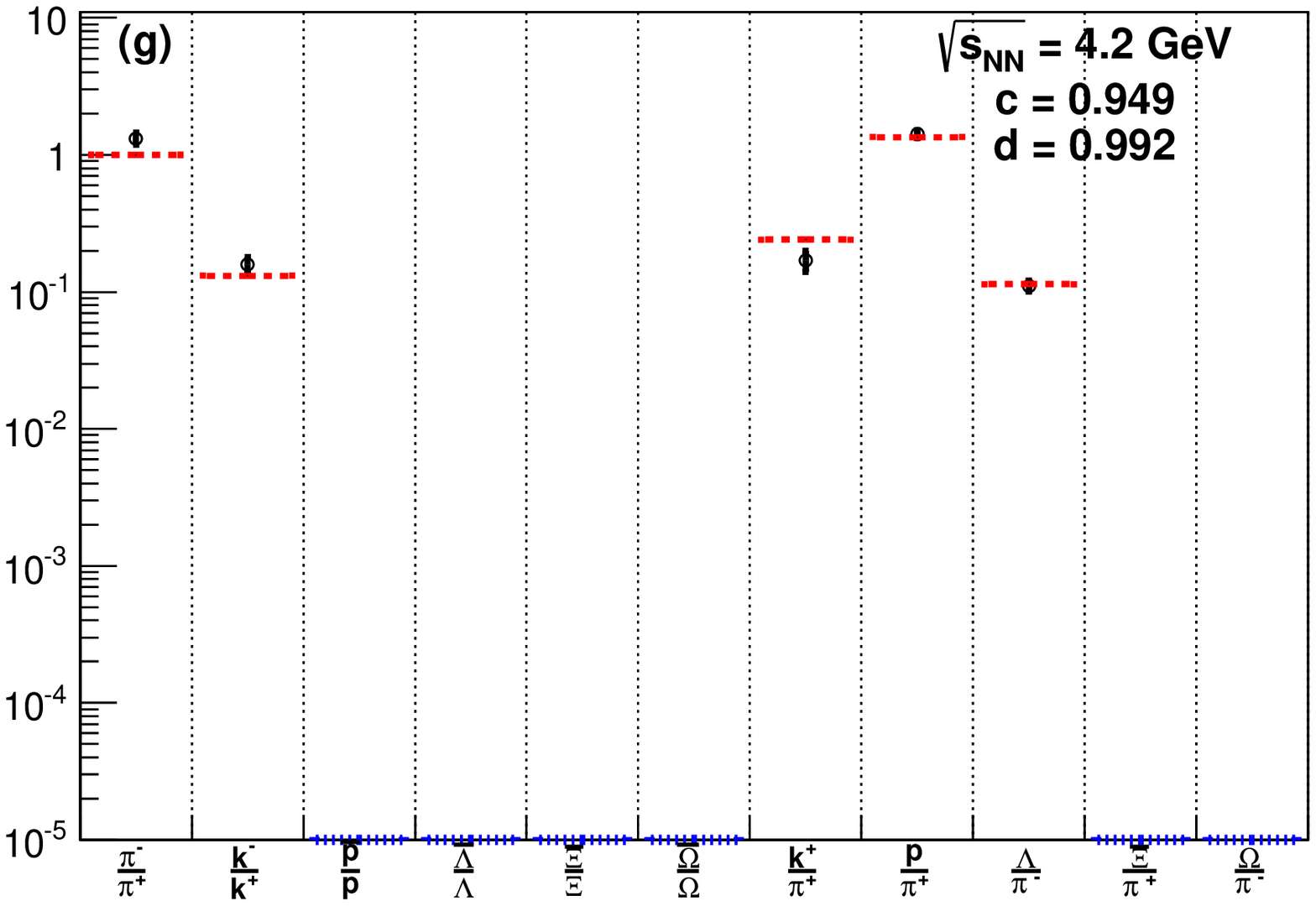}
\includegraphics[width=7.cm]{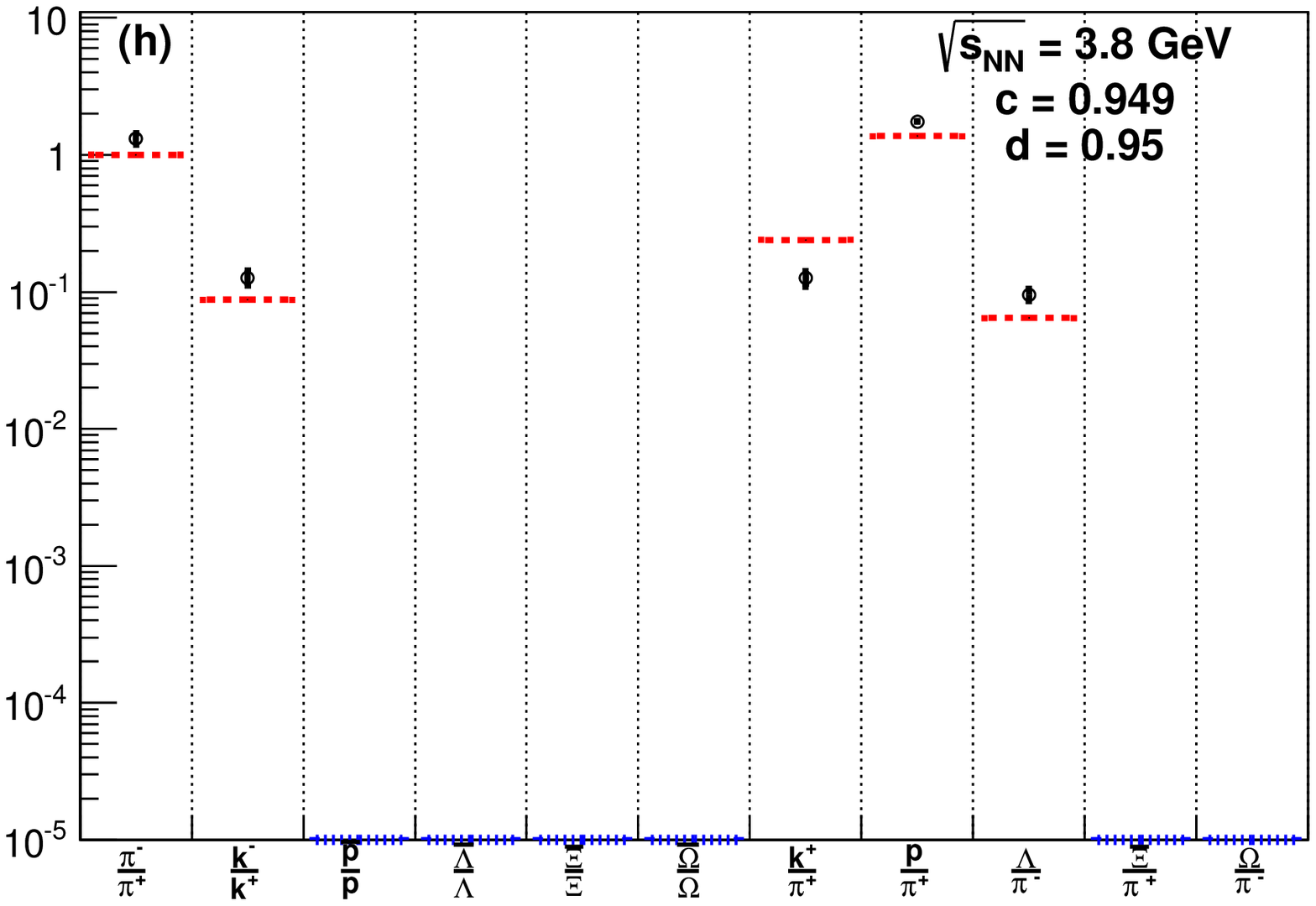}
\caption{The same as in Fig. \ref{RHIC} but at $17$ (a), $12$ (b), $9$ (c), $8$ (d), $6.4~$GeV (e), $5~$ (f), $4.2~$ (g), and $3.8~$ (h). %, and $3.2~$ (i). 
\label{SPS}}
\end{figure}

For the same set of particle ratios, we fit our calculations based on GNS, section \ref{sec:Generic}, Eq. (\ref{FDn1}), to the experimental results. The free parameters are $T_{\mathrm{ch}}$ and  $\mu_{\mathrm{b}}$, besides the scaling exponents ($c$, $d$). There values are deduced where chemical freezeout conditions, such as $s/T^3=7$ \cite{Tawfik:2014eba,Tawfik:2016jzk,Tawfik:2015fda,Tawfik:2013eua,Tawfik:2013dba,Tawfik:2012si,Tawfik:2005qn,Tawfik:2004ss} is fulfilled and minimum $\chi^2$ is obtained, simultaneously. The results are shown in Figs. \ref{Fig1}, \ref{RHIC}, and \ref{SPS}.

\begin{figure}[hbt]
\includegraphics[width=12.cm]{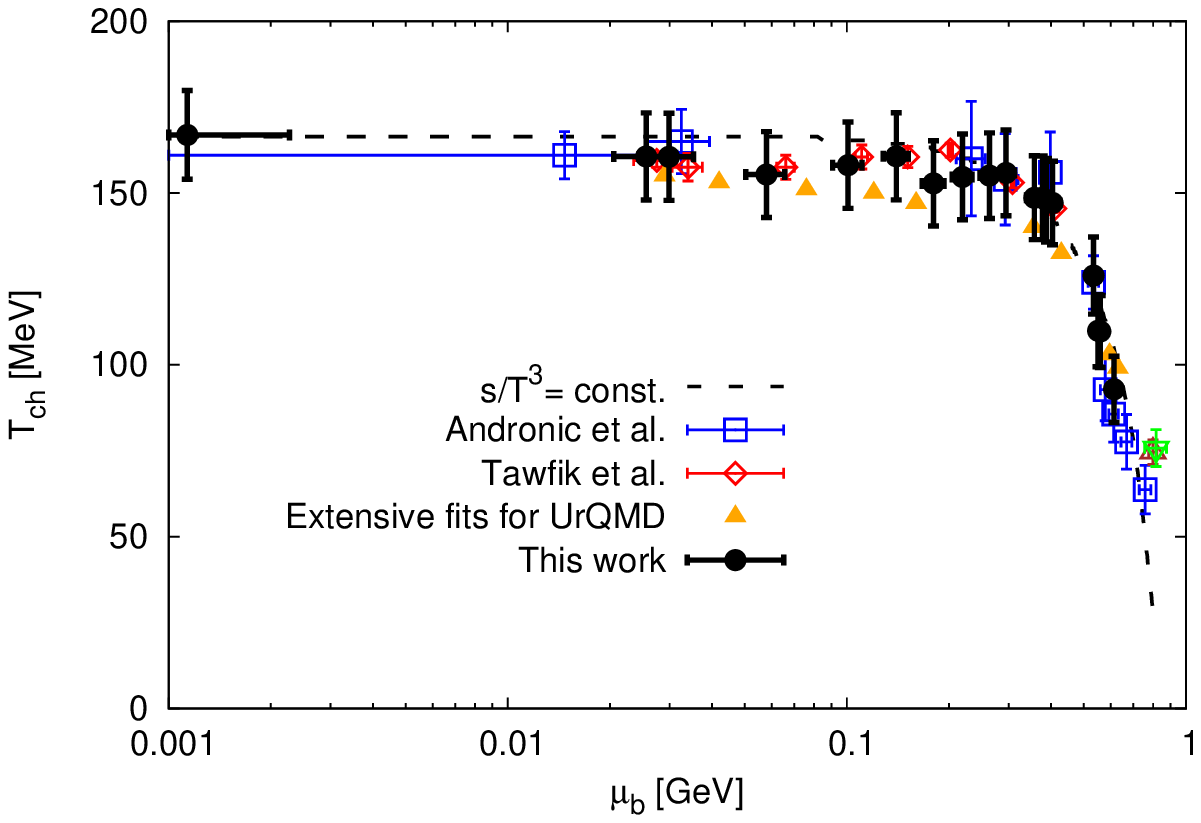}
\caption{The freezeout parameters, temperature $T_{\mathrm{ch}}$ vs. baryon chemical potential $\mu_b$ which haven been deduced from GNS fits of HRG (closed circle) to various particle-ratios, Figs \ref{Fig1}, \ref{RHIC}, and \ref{SPS}. The symbols refer to freezeout parameters deduced from extensive fits: Andronic {\it et al.} \cite{Andronic:2005yp}, Tawfik {\it et al.} \cite{Tawfik:2013bza,Tawfik:2014dha}, and UrQMD \cite{UrQMDppr}.}
\label{Phase_diagram}
\end{figure}

Results on the statistical fits for various particle-ratios measured by ALICE at $2760~$GeV (LHC) [left panel (a)] and by STAR experiment at $200~$GeV (RHIC) [right panel (b)] are depicted in Fig. \ref{Fig1}. The resulting $T_{\mathrm{ch}}$ and  $\mu_{\mathrm{b}}$ are summarized in Tab. \ref{Tab1} and shall be depicted in Fig. \ref{Phase_diagram}. It is obvious that they are very compatible with the ones deduced from BG statistics \cite{Tawfik:2013bza}. But, the resulting exponents ($c$,$d$) for first and second nonextensive property, respectively, differ from BG ($1$, $1$) and Tsallis ($q$, $0$).

So-far, we conclude that the resulting $T_{\mathrm{ch}}$ and $\mu_{\mathrm{b}}$ seem very compatible with the ones assuming extensive statistics (BG). Does this alone mean that the particle productions in heavy-ion collisions is based on extensive processes? The answer is apparently {\it "no"} \cite{Tawfik:2016pwz,Tawfik:2016jol,Tawfik:2017bul}. A key measure would be the reproduction of first-principle calculations, the lattice QCD simulations on thermodynamics, for instance. While HRG with BG statistics excellently models the lattice thermodynamics below the critical temperature and describes well the particle productions in heavy-ion collisions \cite{Karsch:2003vd,Karsch:2003zq,Redlich:2004gp,Tawfik:2004vv,Tawfik:2004sw,Tawfik:2005gk,Tawfik:2005qh}, the Tsallis nonextensivity doesn't do this job. Only, it is conjectured to characterize the kinetic freezeout and fitted well with the measured transverse-momentum distribution \cite{ReFF1,ReFF2}. How to unify these together, especially, when the Tsallis temperature is remarkably smaller than the freezeout temperature? First, let us recall that the Tsallis temperature fails to reproduce the lattice thermodynamics \cite{deppmn0,deppmn1}. The application of the GNS to lattice QCD thermodynamics and particle ratios which measured at different energies of RHIC was already done by one of the authors (AT) in Ref. \cite{1712.04807}. Second, the answer has been delivered, couple years ago. In previous sections, we discussed on this that, the Tsallis resulting temperature should be extrapolated to the freezeout one, $T_{\mathrm{ch}}$. The present work proposes another answer.

Figure \ref{Fig1} compares experimental results on various particle-ratios (circles), $\mathrm{\pi}^-/\mathrm{\pi}^+$, $\mathrm{K}^-/\mathrm{K}^+$, $\bar{\mathrm{p}}/\mathrm{p}$, $\bar{\mathrm{\Lambda}}/\mathrm{\Lambda}$, $\bar{\mathrm{\Omega}}/\mathrm{\Omega}$, $\bar{\mathrm{\Xi}}/\mathrm{\Xi}$, $\mathrm{K}^-/\mathrm{\pi}^-$, $\mathrm{K}^+/\mathrm{\pi}^+$, $\bar{\mathrm{p}}/\mathrm{\pi}^-$, $\mathrm{p}/\mathrm{\pi}^+$, $\mathrm{\Lambda}/\mathrm{\pi}^-$, $\mathrm{\Omega}/\mathrm{\pi}^-$, and $\bar{\mathrm{\Xi}}/\mathrm{\pi}^+$ measured by ALICE- and STAR-experiment at energies $2760~$GeV and $200~$GeV, left and right panel, respectively, are statistically fitted by means of the HRG model in which GNS is implemented. In our HRG calculations, all possible decay channels yielding the particles of interest and the related branching ratios are taken into account, Eq. (\ref{eq:nn1}). For the decay channels with not-yet-measured probabilities, the rules given in Ref. \cite{Andronic:2005yp} have been applied. But no finite-size correction was applied \cite{r6}. We sum up contributions from all hadron resonances listed in recent particle data group compilation with masses $\leq 2~$GeV. This refers to $388$ different states of mesons and baryons besides their anti-particles. For further details, interested readers can consult  Ref. \cite{Tawfik:2014eba}. The number density can be derived from the partition function and accordingly the particle ratios, Eq. (\ref{FDn1}), can be determined. Other GNS fits are illustrated in Fig. \ref{RHIC} (RHIC) and Fig. \ref{SPS} (SPS).

In Fig. \ref{Phase_diagram}, the freezeout parameters, $T_{\mathrm{ch}}$ and $\mu_{\mathrm{b}}$ (closed circles) as deduced from the statistical fits of the generic-nonextensive HRG are compared with the ones from the extensive HRG (dashed line). At a finite chemical potential, the freezeout temperature in both cases is determined at constant $s/T^3$, where $s$ is the entropy density \cite{Tawfik:2014eba}. Apparently, both approaches are very compatible with each other. In other words, there is almost no difference between the resulting {\it "physical temperature"} and the {\it "temperature"} ($T_{\mathrm{ch}}$). We also compare with other freezeout parameters (symbols) determined in different phenomenologies; Andronic {\it et al.} \cite{Andronic:2005yp}, Tawfik {\it et al.}  \cite{Tawfik:2013bza,Tawfik:2014dha}, and UrQMD \cite{UrQMDppr}. With the latter we mean various simulations by hybrid UrQMD version $3.4$ at varying chemical potentials. At each value of the chemical potential, the simulated particle ratios are fitted by means of extensive HRG. In other words, UrQMD simulations - in this case - are taken as experimental results. To this end, it was shown that the UrQMD simulations agree well with measured particle-ratios. Further details can be taken from Ref. \cite{UrQMDppr}.

The resulting freezeout temperature and chemical potential are almost the same when extensive and GNS fits are applied. As discussed in earlier sections, it was widely believed that the nonextensivity in the particle production should be accompanied by a radical change in intensive or extensive thermodynamic quantities, such as the internal energy and the temperature. Therefore, the resulting temperature, for instance, was taken as a so-called {\it "physical"} one, which afterwards should be extrapolated to the freezeout temperature. The short-cuts of these assumptions shall be discussed in a forthcoming work. To remain within the scope of the present word, we report on nonextensive-statistical fit of various particle-ratios, where the intensive and extensive thermodynamic properties of the strongly interacting system remain almost unchanged, while the nonextensitivity is defined by the equivalence classes $(c, d)$. Their values are found close to unity, Tab. \ref{Tab1}. This refers to a generic nonextensivity, which is apparently not of Tsallis-type. The latter is characterized by ($q, 0$) with $q>1$.

It should be remarked that the excellent fits of various particle-ratios to the GNS statistical approach shouldn't be interpreted due to adding extra parameters ($c, d$). This should be related to rightly implementing generic statistical approach. Should the statistical nature of producing various particle-ratios were really Tsallis nonextensivity, the added extra parameters should be ($q, 0$). Should this were extensive, ($1, 1$) should be resulted in. On the other hand, the success of the GNS statistical approach in describing the lattice thermodynamics at ($1, 1$) confirms the capability of  the proposed GNS to determine whether the system of interest has an extensive or a nonextensive statistical nature and apparently allows its implementation in other ambiguous processes.

\subsection{Equivalence classes $(c, d)$}
\label{sec:resultingcd}

Table \ref{Tab1} summarizes the freezeout parameters ($T_{\mathrm{ch}}$ and $\mu_{\mathrm{b}}$) and the scaling exponents ($c$, $d$) as determined from the statistical fits of HRG with the proposed generic nonextensivity, GNS, to various experimental results. Both freezeout parameters are graphically illustrated in Fig. \ref{Phase_diagram} and compared with other results, section \ref{sec:fittingPR}.

The resulting classes ($c$, $d$) differ from unity referring to Lambert-$W$ exponentials characterizing the entropic equivalence classes. The generalized logarithm and exponential  functions were given in Eq. (\ref{eq:genLog}) and (\ref{eq:epsln}), respectively. This means that the particle productions has a non-BG statistical nature. Over the whole energy-range, the entropic equivalence classes can be expressed as Lambert-$W$ exponentials, which is generic. Both  resulting classes are positive but less than unity, except at LHC energy. We observe a general trend that the exponent $c$ is closer to unity than $d$. At $\sqrt{s}\approx 20~$GeV, $d$ reaches minimum, while $c$ nearly maximum values. The differences between the two exponents becomes great at low-SPS energies.

The maximum error in the exponent is only $10\%$. Taking into account a wide energy range, during which relevant microscopic physics radically changes, this is excellently precise.

Let us take $d$ as resulted, i.e. $d>0$, and assume that the resulting $c$ can be approximated to unity. Accordingly, the resulting entropy is characterized by stretched exponential,
\bea
\varepsilon_{(1,d,r)}(x) &=& \exp\left\{-d\, r \left[\left(1-\frac{x}{r}\right)^{1/d} - 1\right]\right\}.
\eea
Mathematically, this function can be interpreted as a fractional power law. A {\it critical} exponent is inserted into an {\it ordinary} exponential function. Accordingly, in a disordered system, for instance, the particle production, this type of entropy characterizes a delayed relaxation.

\section{Conclusions and final remarks}
\label{sec:cncl}

In the present work, we have introduced a systematic study for (non)extensive properties characterizing relativistic heavy-ion collisions, which are assumed to hadronize and then i.e. forming hadrons or particles. With the nowadays detector technologies, the latter - in turn - can very precisely be detected. In doing this, we imply GNS on the well-known statistical-thermal models, such as the HRG model. It is assumed that the degree of (non)extensivity can be determined by the phase space characterizing the (dis)ordered system of interest. In other words, the proposed approach is able to determine whether the system of interest has extensive or nonextensive properties. Furthermore, more details about the nonextensivity can be also determined.

The particle ratios at energies ranging between $3.8$ and $2760~$GeV are best reproduced by GNS, where the equivalence classes ($c$,$d$) range between $\sim0.9$ and $\sim1$. This leads to a crucial conclusion that the statistics describing the particle production at a wide range of beam energies remarkably differ from extensive BG ($1, 1$) or nonextensive Tsallis ($q, 0$). It is certainly interesting to see how the economic approach (extensive BG), as it deals with smallest free parameters, works quite well. In fact, this is the main conclusion of this present work, either generic statistical approach with $(c,d)=(\sim1, \sim1)$ and the extensive BG approach reproduces well various particle ratios in a wide range of energies. Therefore, when comparing our results with the $q$-entropies, we can propose to refute all previous relevant works in the literature. Such a simplification will strengthen the discussion. The advantage of the present work is now simply to reassure that the particle ratios have extensive statistical nature but to illustrate that the proposed generic approach indeed manifests the degree of (non)extensivity depending on the resulting equivalence classes $(c, d)$. In a previous work, we have examined this with the lattice QCD thermodynamics, which is per definition assumes additivity (extensivity). Now, we present another examination with the particle ratios produced in heavy-ion collisions at various beam energies, which were ambiguously analyzed by thermal models assuming an {\it ac hoc} extensive statistics.

Furthermore, the resulting equivalence classes imply that the system of interest can best be described by an extended exponential entropy, which basically differs from the well-know extensive BG and from the well-know nonextensive Tsallis entropy. Both are very special cases defined by specific values of the equivalence classes $(1, 1)$ and $(q, 0)$, respectively. We conclude that the particle production, in terms of the produced particle-ratios, in relativistic heavy-ion collisions seems to be originated from a dynamically disordered system. 

Our results propose a plausible interpretation why the resulting Tsallis freezeout-temperature, which is called {\it "physical temperature"}, differs from the resulting BG freezeout-temperature, which is called the {\it "temperature"}. We believe that the insist to imply Tsallis-type nonextensivity to the high-energy particle productions seems to be accompanied with a high price; a radical change in intensive and/or extensive thermodynamic quantities such as temperature and internal energy.  Our results propose that the resulting freezeout parameters are compatible with the ones obtained when BG statistics is implied. Accordingly, the well-known freezeout phase-diagram  excellently agrees with the recent lattice predictions and with the freezeout phase-diagram based on extensive BG statistics. The latter is very obvious, especially when comparing the equivalence classes obtained in our calculations and the ones characterizing BG. We conclude that the nonextensivity characterizing the high-energy particle productions is solely through two critical exponents defining equivalence classes $(c, d)$.

A few final remark on the excellent fits reported in this paper is now in order. First, as discussed, that the GNS partition function, Eq. (\ref{eq1}), was implemented, in which an additional pair of free parameters, ($c,d$), was added, shouldn't be thought as an explanation based on statistical precision, which likely increases with increasing the number of the free parameters. The other pair of free parameters, $T_{\mathrm{ch}}$ and $\mu_{\mathrm{b}}$, agrees well with the extensive fits, while the resulting ($c,d$) can be approximated to unity, i.e. almost extensive. Thus, the GNS partition function, Eq. (\ref{eq1}), with its four free parameters is not just a statistical precision. It manifests a generic statistical approach that seems to characterize various aspects for the particle productions. Because of its generic nature, both special cases, BG and Tsallis, are also included. Its equivalence classes $(c, d)$ define the degree of (non)extensivity.

It is obvious that the statistical physics approaches to the high-energy phenomena have to be understood as an economic description of the phenomena of interest in terms of a small number of variables including the thermodynamic variables. This is true as long as the statistical nature hasn't be changed. Tsallis statistics is not only introducing an additional parameter, $q$, but also an special concept of nonextensivity. In Euclidean field theory or lattice field theory, change of exponential factor implies change of quantum theory, which does not seem to be possible as the extensivity is fundamentally imposed. The present paper presents an interpretation to the high-energy phenomena, the production of various particle ratios, through suggesting changes of the foundations of statistical physics.

%=========================================================================

%\appendix
%\section{Fits for other particle ratios at various beam-energies}

\begin{table}[tb]
\begin{center}
\begin{tabular}{||c||c|c|c|c||c||}\hline \hline
  $\sqrt{s_{\mathrm{NN}}}$ (GeV) & $T_{\mathrm{ch}}$ (MeV) &$\mu_{\mathrm{b}}$ (MeV)& c & d & $\chi^2/{\mathrm{dof}}$ \\ \hline \hline
  $ 2760$ &166.93  &1.134  &1.009   &1.031   &7.3     \\ \hline
  $ 200$  &160.65  &25.578 &0.969   &0.9475  &2.24    \\ \hline
  $ 130$  &160.54  &29.862 &0.971   &0.9495  &1.65   \\ \hline
  $ 62.4$ &155.4   &57.96  &0.965   &0.982   &2.78    \\ \hline
  $ 39  $ &158.13  &100.8  &0.99    &0.963   &12.17   \\ \hline
  $ 27 $  &160.71  &139.86 &0.991   &0.93    &19.98    \\ \hline
  $ 19.6$ &152.81  &180.18 &0.988   &0.902   &2.1    \\ \hline
  $ 17  $ &154.71  &219.24 &0.99    &0.895   &20.15   \\ \hline
  $ 12  $ &155.08  &263.34 &0.991   &0.908   &2.11    \\ \hline
  $ 11.5$ &155.84  &294.84 &0.989   &0.931   &7.1    \\ \hline
  $ 9   $ &148.64  &357.21 &0.991   &0.916   &1.33      \\ \hline
  $8    $ &148.63  &377.37 &0.972   &0.939   &4.99    \\ \hline
  $7.7  $ &147.88  &388.08 &0.991   &0.939   &11.02    \\ \hline
  $6.4  $ &147.11  &404.46 &0.97    &0.9573  &12.71    \\ \hline
  $5    $ &126.    &534.24 &0.952   &0.994   &3.01    \\ \hline
  $4.2  $ &109.94  &550.62 &0.949   &0.992   &14.18    \\ \hline
  $3.8  $ &109.73  &559.44 &0.949   &0.95    &67.61    \\ \hline \hline
%  $3.2  $ &92.88   &613.62 &0.942   &0.99    &141.58    \\ \hline
 \end{tabular}
 \end{center}
 \caption{The freezeout parameters ($T_{\mathrm{ch}}$ and $\mu_{\mathrm{b}}$) and the scaling exponents  ($c$, $d$) as determined from the proposed GNS fits to various experimental results. 
 \label{Tab1}}
 \end{table}

\begin{thebibliography}{99}
%=========================================================================

\bibitem{Tawfik:2014eba} A Tawfik
%{\it "Equilibrium statistical-thermal models in high-energy physics"}, 
{\it Int J Mod Phys A} {\bf 29} 1430021 (2014)  %1410.0372 [hep-ph]

\bibitem{Tawfik:2010aq} A Tawfik 
%{\it "Dynamical fluctuations in baryon-meson ratios"}, 
{\it J Phys G} {\bf 40} 055109 (2013) %1007.4585

\bibitem{Hagedorn1965} R Hagedorn
%{\it "Statistical thermodynamics of strong interactions at high energies"}, 
{\it Nuovo Cimento Suppl} {\bf 3} 147 (1965)

\bibitem{Tawfik:2010uh} A Tawfik 
%{\it "Phase space and dynamical fluctuations of kaon-to-pion ratios"}, 
{\it Prog Theor Phys} {\bf 126} 279 (2011) %1007.4074 [hep-ph]

\bibitem{ReFF1} C Beck
%{\it "Non-extensive statistical mechanics and particle spectra in elementary interactions"}, 
{\it Physica A} {\bf 286} 164 (2000)

\bibitem{ReFF2} I Bediaga E M F Curado J M de Miranda 
%{\it "A nonextensive thermodynamical equilibrium approach in $e^+e^- \rightarrow$ hadrons"}, 
{\it Physica A} {\bf 286} 156 (2000)

\bibitem{Tsallis:1987eu} C Tsallis {\it J Stat Phys} {\bf 52}  479 (1988)

\bibitem{Prato:1999jj} D Prato and C Tsallis {\it Phys Rev E} {\bf 60} 2398 (1999)

\bibitem{Tawfik:2016pwz} A  Tawfik 
%{\it "Axiomatic nonextensive statistics at NICA energies"}, 
{\it Eur Phys J A} {\bf 52} 253 (2016) %1607.01264 [nucl-th].

\bibitem{Tawfik:2016jol} A Tawfik 
%{\it ''Baryon-to-pion ratios within generic (non)extensive statistics''}, 
Proceedings, 38th International Conference on High Energy Physics (ICHEP 2016): Chicago, IL, USA, August 3-10, (2016)
%1701.05423 [nucl-th]

\bibitem{Tawfik:2017bul}  A Tawfik H Yassin E R Abo Elyazeed
%{\it ''On thermodynamic self-consistency of generic axiomatic-nonextensive statistics''}, 
{\it Chin Phys C} {\bf 41} 053107  (2017)
%1701.04697 [nucl-th]

\bibitem{Cleymans:2011in} J Cleymans and D Worku {\it J Phys G} {\bf 39} 025006 (2012)

\bibitem{Azmi:2014dwa} M D Azmi and J Cleymans {\it J Phys G} {\bf 41} 065001 (2014)

\bibitem{AbeRef} S Abe
%{\it "Temperature of nonextensive systems: Tsallis entropy as Clausius entropy"},
{\it Physica A} {\bf 368} 430 (2006)

\bibitem{Refeee1} S Abe
%{\it ''Heat and entropy in nonextensive thermodynamics: transmutation from Tsallis theory to Renyi-entropy-based theory''}, 
{\it Physica A} {\bf 300} 417 (2001)

\bibitem{Refeee2} S Abe
%{\it ''General pseudoadditivity of composable entropy prescribed by the existence of equilibrium''}, 
{\it Phys Rev E} {\bf 63} 061105 (2001)

\bibitem{Refeee3} T S Biro  P Van 
%{\it '' Zeroth law compatibility of nonadditive thermodynamics''}, 
{\it Phys Rev E} {\bf 83} 061147 (2011)

\bibitem{Thurner1} R Hanel S Thurner
%{\it "A comprehensive classification of complex statistical systems and an axiomatic derivation of their entropy and distribution functions"},
{\it Europhys Lett} {\bf 93} 20006 (2011)

\bibitem{Thurner2} R Hanel  S Thurner
%{\it "When do generalized entropies apply? How phase space volume determines entropy"},
{\it Europhys Lett} {\bf 96} 50003 (2011)

%\bibitem{Thurner3} R. Hanel and S. Thurner, {\it "A comprehensive classification of complex statistical systems and an ab initio derivation of their entropy and distribution functions"}, Europhys. Lett. {\bf 93} 20006, (2011).

\bibitem{deppmn0} A Deppman
%{\it "Thermodynamics with fractal structure, Tsallis statistics and hadrons"},
{\it Phys Rev D} {\bf 93} 054001 (2016) % arXiv:1601.02400 [hep-ph]

\bibitem{deppmn1} A Deppman 
%{\it "Self-consistency in non-extensive thermodynamics of highly excited hadronic states"},
{\it Physica A} {\bf 391} 6380 (2012) %arXiv:1205.0455 [math-ph]

\bibitem{deppmn2} I Sena  A Deppman
%{\it "Systematic analysis of transverse momentum distribution and non-extensive thermodynamics theory"},
{\it AIP Conf Proc} {\bf 1520} 172 (2013) % arXiv:1208.2952 [hep-ph]

\bibitem{tamas1} T S Biro G G Barnafoldi P Van
%{\it "Quark-gluon plasma connected to finite heat bath"},
{\it Eur Phys J A} {\bf 49} 110 (2013) %arXiv:1208.2533 [hep-ph]

\bibitem{Deppmann2015} E Megias D P Menezes A Deppman
%{\it "Non extensive thermodynamics for hadronic matter with finite chemical potentials"}, 
{\it Physica A} {\bf 421} 15 (2015)

\bibitem{Ding:2015ona} H-T Ding F Karsch S Mukherjee
%{\it "Thermodynamics of strong-interaction matter from Lattice QCD"},
{\it Int J Mod Phys E} {\bf 24} 1530007 (2015)
%1504.05274 [hep-lat]

\bibitem{Tawfik:2013bza} A Tawfik  E Abbas
%{\it "Thermal description of particle production in Au-Au collisions at STAR energies"},
{\it Phys Part Nucl Lett} {\bf 12} 521 (2015) %1311.7508 [nucl-th].


\bibitem{Tawfik:2014dha}  A Tawfik et al.
%{\it "Degree of chemical non-equilibrium in central Au-Au collisions at RHIC energies"},
{\it Int J Mod Phys E} {\bf 24} 1550067 (2015)  %1401.5715 [nucl-th]

\bibitem{Karsch:2003vd}  F Karsch K Redlich A Tawfik
%{\it "Hadron resonance mass spectrum and lattice QCD thermodynamics"},
{\it  Eur Phys J C} {\bf 29} 549 (2003)
%hep-ph/0303108

\bibitem{Karsch:2003zq}  F Karsch K Redlich A Tawfik
%{\it "Thermodynamics at nonzero baryon number density: A comparison of lattice and hadron resonance gas model calculations"},
{\it Phys Lett B} {\bf 571} 67 (2003)
%hep-ph/0306208

\bibitem{Redlich:2004gp} K Redlich F Karsch A Tawfik
%{\it "Heavy ion collisions and lattice QCD at finite baryon density"},
{\it J Phys G} {\bf 30} S1271 (2004)
%nucl-th/0404009

\bibitem{Tawfik:2004vv} A Tawfik
%{\it "The Influence of strange quarks on QCD phase diagram and chemical freeze-out: Results from the hadron resonance gas model"},
{\it J Phys G} {\bf 31} S1105 (2005)
%hep-ph/0410329

\bibitem{Tawfik:2004sw} A Tawfik
%{\it "QCD phase diagram: A comparison of lattice and hadron resonance gas model calculations"},
{\it Phys Rev D} {\bf 71} 054502  (2005)
%hep-ph/0412336

\bibitem{Tawfik:2005gk} A Tawfik
%{\it "Particle ratios in heavy-ion collisions"},
{\it Fizika B} {\bf 18} 141 (2009)
%hep-ph/0508244

\bibitem{Tawfik:2005qh} A Tawfik D Toublan
%{\it "Quark-antiquark condensates in the hadronic phase"},
{\it Phys Lett B} {\bf 623} 48 (2005)
%hep-ph/0505152

\bibitem{Tawfik:2015} A  Tawfik 
%{\it "Lattice QCD thermodynamics and RHIC-BES particle production within generic nonextensive statistics"}, 
submitted to PEPAN Lett

\bibitem{RAFELSKI} J Letessier J Rafelski 
{\it "Hadrons and Quark-Gluon Plasma"} (UK:Cambridge University Press) T Ericson P V Landshoff (2004)

\bibitem{Ahle:2000wq} L Ahle  et al. (E866 Collaboration and E917 Collaboration)
%{\it "An Excitation function of K- and K+ production in Au + Au reactions at the AGS"}, 
{\it Phys Lett B} {\bf 490} 53 (2000) %nucl-ex/0008010
    
\bibitem{Klay:2001tf} J L Klay  et al. (E895 Collaboration)
%{\it "Longitudinal flow from 2-A-GeV to 8-A-GeV Au+Au collisions at the Brookhaven AGS"}, 
{\it Phys Rev Lett} {\bf 88} 102301 (2002) % nucl-ex/0111006
    
\bibitem{Ahle:1999uy} L Ahle et al. (E866 Collaboration and E917 Collaboration)
%{\it "Excitation function of K+ and pi+ production in Au + Au reactions at 2/A-GeV to 10/A-GeV"}, 
{\it Phys Lett B} {\bf 476} 1 (2000) % nucl-ex/9910008

\bibitem{Andronic:2005yp} A Andronic P Braun-Munzinger J Stachel
%{\it "Hadron production in central nucleus-nucleus collisions at chemical freeze-out"},
{\it Nucl Phys A} {\bf 772} 167 (2006) %nucl-th/0511071v3.

\bibitem{Tawfik:2016jzk} A Tawfik M Y El-Bakry D M Habashy M T Mohamed E Abbas
%{\it "Possible interrelations among chemical freeze-out conditions"},
{\it Int J Mod Phys E} {\bf 25} 1650018  (2016)
%1603.07085 [hep-ph]

\bibitem{Tawfik:2015fda} A Tawfik H Yassin  E R. Abo Elyazeed
%{\it "Chemical freeze-out in Hawking-Unruh radiation and quark-hadron transition"},
{\it Phys Rev D} {\bf 92} 085002 (2015)
%1510.02117 [hep-ph]

\bibitem{Tawfik:2013eua} A Tawfik
%{\it "Constant Trace Anomaly as a Universal Condition for the Chemical Freeze-Out"},
{\it Phys Rev C} {\bf 88} 035203 (2013)
%1308.1712 [hep-ph]

\bibitem{Tawfik:2013dba} A Tawfik
%{\it "Chemical Freeze-Out and Higher Order Multiplicity Moments"},
{\it Nucl Phys A} {\bf 922} 225 (2014)
%1306.1025 [hep-ph]

\bibitem{Tawfik:2012si} A Tawfik
%{\it "On the higher moments of particle multiplicity, chemical freeze-out and QCD critical endpoint"},
{\it Adv High Energy Phys} {\bf 2013} 574871 (2013)
%1205.1761 [hep-ph]

\bibitem{Tawfik:2005qn} A Tawfik
%{\it "A Universal description for the freezeout parameters in heavy-ion collisions"},
{\it Nucl Phys A} {\bf 764} 387 (2006)
%hep-ph/0508273

\bibitem{Tawfik:2004ss} A Tawfik
%{\it "On the conditions driving the chemical freeze-out"},
{\it Europhys Lett} {\bf 75} 420 (2006)
%hep-ph/0410392

\bibitem{1712.04807} A Tawfik {\it PEPAN Lett} {\bf 15} 1 (2018)

\bibitem{r6} P Braun-Munzinger J Stachel J P Wessels N Xu 
%{\it "Thermal and hadrochemical equilibration in nucleus-nucleus collisions at the SPS"},
{\it Phys Lett B} {\bf 365} 1 (1996)

\bibitem{UrQMDppr}  A Tawfik L I Abou-Salem A G Shalaby M Hanafy A Sorin O Rogachevsky W Scheinast
%{\it "Particle production and chemical freezeout from the hybrid UrQMD approach at NICA energies"},
{\it Eur Phys J A} {\bf 52} 324 (2016) 
%1609.08423 [nucl-th].

\end{thebibliography}
\end{document}